\documentclass[structabstract]{aa}  
\usepackage{graphicx}
\usepackage{txfonts}
\usepackage{natbib,twoopt}
\usepackage{epsfig}
\usepackage{graphicx}
\usepackage{tabularx}
\usepackage{enumerate}
\usepackage{pdflscape}
\usepackage{bm}
\usepackage{color}

\begin{document}

   \title{Fossil group origins}

   \subtitle{V. The dependence of the luminosity function on the magnitude gap}

    \author{S. Zarattini \inst{1, 2,3}, J.A.L. Aguerri\inst{1,2}, R. S\'anchez-Janssen\inst{4}, R. Barrena \inst{1, 2}, W. Boschin\inst{1,5}, C. del Burgo\inst{6}, \\
        N. Castro-Rodriguez\inst{1, 2}, E. M. Corsini\inst{3}, E. D'Onghia\inst{7}, M. Girardi \inst{8,9}, J. Iglesias-P\'aramo\inst{10,11}, A. Kundert\inst{7}, \\
        J. M\'endez-Abreu\inst{12}, J.M. Vilchez\inst{10}}
          
                   %and

     \institute{Instituto de Astrofísica de Canarias, calle Via L\'actea S/N, E-38205, La Laguna, Tenerife, Spain
     \and Universidad de La Laguna, Dept. Astrof\'isica, E-38206, La Laguna, Tenerife, Spain
     \and Dipartimento di Fisica e Astronomia "G. Galilei", Universit\`a di Padova, vicolo dell'Osservatorio 3, 35122 Padova, Italy
     \and NRC Herzberg Institute of Astrophysics, 5071 West Saanich Road, Victoria, BC, V9E 2E7, Canada
     \and Fundación Galileo Galilei - INAF, Rambla José Ana Fernández Pérez 7, E-38712, Breña Baja, La Palma, Spain
     \and Instituto Nacional de Astrofísica, \'Optica y Electrónica (INAOE), Luis Enrique Erro 1, Sta. Ma. Tonantzintla, Puebla, Mexico
     \and Astronomy Department, University of Wisconsin, 475 Charter St., Madison, WI 53706, USA\footnote{Alfred P. Sloan Fellow}
     \and Dipartimento di Fisica-Sezione Astronomia, Università degli Studi di Trieste, Via Tiepolo 11, I-34143, Trieste, Italy
     \and INAF - Osservatorio Astronomico di Trieste, via Tiepolo 11, I-34143, Trieste, Italy
     \and Instituto de Astrofísica de Andalucía – C.S.I.C., E-18008, Granada, Spain
     \and Estaci\'{o}n Experimental de Zonas Aridas (CSIC), Ctra. de Sacramento s/n, La Ca\~{n}ada, Almer\'{\i}a, Spain
     \and School of Physics and Astronomy, University of St Andrews, North Haugh, St Andrews, KY16 9SS, UK
      \\email{stefano@iac.es}}

   \date{Accepted 07/05/2015}

 %\abstract{}{}{}{}{} 
% 5 {} token are mandatory
 
  \abstract
   {In nature we observe galaxy aggregations that span a wide range of magnitude gaps between the two first-ranked galaxies of a system ($\Delta m_{12}$). Thus, there are systems with gaps close to zero (e.g., the Coma cluster), and at the other extreme of the distribution, the largest gaps are found among the so-called fossil systems. The observed distribution of magnitude gaps is thought to be a consequence of the orbital decay of $M^\ast$ galaxies in massive halos and the associated growth of the central object. As a result, to first order the amplitude of this gap is a good statistical proxy for the dynamical age of a system of galaxies. Fossil and non-fossil systems could therefore have different galaxy populations that should be reflected in their luminosity functions.}
   {In this work we study, for the first time, the dependence of the luminosity function parameters on $\Delta m_{12}$ using data obtained by the fossil group origins (FOGO) project.}
   {We constructed a hybrid luminosity function for 102 groups and clusters at $z \le 0.25$ using both photometric data from the SDSS-DR7 and  redshifts from the DR7 and the FOGO surveys. The latter consists of $\sim$1200 new redshifts in 34 fossil system candidates. We stacked all the individual luminosity functions, dividing them into bins of $\Delta m_{12}$, and studied their best-fit Schechter parameters. We additionally computed a ``relative'' luminosity function, expressed as a function of the central galaxy luminosity, which boosts our capacity to detect differences---especially at the bright end.}
   {We find trends as a function of $\Delta m_{12}$ at both the bright and faint ends of the luminosity function. In particular, at the bright end, the larger the magnitude gap, the fainter the characteristic magnitude $M^\ast$. The characteristic luminosity in systems with negligible gaps is more than a factor three brighter than in fossil-like ones. Remarkably, we also find differences at the faint end. In this region, the larger the gap, the flatter the faint-end slope $\alpha$.}
   {The differences found at the bright end support a dissipationless, dynamical friction-driven merging model for the growth of the central galaxy in group- and cluster-sized halos. The differences in the faint end cannot be explained by this mechanism. Other processes---such as enhanced tidal disruption due to early infall and/or prevalence of eccentric orbits---may play a role. However, a larger sample of systems with $\Delta m_{12} > 1.5$ is needed to establish the differences at the faint end.}
  % \keywords{}
        \authorrunning{Zarattini et al.}
        \titlerunning{Fossil Groups Origins V}
\maketitle

%%%%%%%%%%%%%%%%%%%%%%%%%%%%%%%%%%
%%%%%%%%%%%%% SECTION 1%%% %%%%%%%%%%%%
%%%%%%%%%%%%%%%%%%%%%%%%%%%%%%%%%%
\section{Introduction}
\label{intro}

The existence of fossil galaxy groups was proposed for the first time by \citet{Ponman1994}. In that work, it was suggested that the isolated elliptical galaxy RX J1340.6+4018 was probably an evolved compact group of galaxies. They claimed that those galaxies that were close to the center of the system could have merged in a single elliptical galaxy. This is why they are called "fossil groups" (FGs). The most accepted observational definition for this kind of object was proposed by \citet{Jones2003}. They defined a system of galaxies as a fossil if it presents a magnitude gap of at least two magnitudes between the two brightest member galaxies ($\Delta m_{12} \ge 2$) in the $r$-band within half of its virial radius and if the central galaxy is surrounded by an extended X-ray halo of $L_{X,{\rm bol}} > 10^{42} h_{50}^{-2}$ erg s$^{-1}$. The latter criterion was adopted to distinguish large isolated galaxies from group-sized systems, but it is a lower limit, so it does not exclude the existence of "fossil clusters" \citep[as proposed by ][]{Cypriano2006,MendesdeOliveira2006,Zarattini2014}. For this reason, we refer to fossil systems, but we prefer to maintain the classical notation of FGs, which is usually accepted in the literature, as we did in the other papers of this series.

Different observational properties of FGs were studied. The properties of the hot intracluster component were analyzed using scaling relations that include some X-ray properties of the system. The $L_X - T_X$ relation is generally similar to that of normal clusters \citep[see][]{Khosroshahi2007, Harrison2012}, whereas differences have been found in scaling relations that combine both optical and X-ray properties, such as the $L_X - L_{\rm opt}$, $L_X - \sigma_{\rm v}$, and $T_X - \sigma_{\rm v}$ relations. In fact, some works exploring these relations suggest that fossil systems are brighter in the X-ray range (or fainter in the optical range) than normal groups and clusters \citep[for example, see][]{Proctor2011}. In contrast, \citet{Khosroshahi2014} find that fossils are underluminous in the X-ray range. However, these differences can be attributed to observational biases \citep[see][]{Voevodkin2010,Harrison2012}. Recently, \citet{Girardi2014} have analyzed a sample of 15 spectroscopically confirmed FGs, finding no significant differences in the $L_X-L_{\rm opt}$ relation. Particular attention was paid to the homogeneity of the data set and the analysis process in this work.

The brightest group galaxies (BGGs) of FGs are considered the most massive galaxies in the Universe, and as such they have been studied well in the literature. For example, \citet{Harrison2012} show that both the absolute magnitude of the BGGs and the fraction of light contained in them are correlated with the magnitude gap, results that we have recently confirmed in \citet{Zarattini2014}. Moreover, their luminosity is correlated with the system velocity dispersion \citep{Khosroshahi2006}. Observations of the BGGs isophotal shape are not conclusive. \citet[][]{Khosroshahi2006} show that these objects present disky isophotes in the central part, whereas both isolated ellipticals and central ellipticals in clusters show boxy isophotes. In contrast, \citet[][]{LaBarbera2009} and \citet{Mendez-Abreu2012} find no differences in this sense between fossil and non-fossil systems. In addition, the size-luminosity relation, the fundamental plane, and the Faber-Jackson relation are similar for fossil and non-fossil central galaxies \citep{Mendez-Abreu2012}. Recent studies of the stellar population of BGGs seem to indicate that their age, metallicity, and $\alpha$ enhancement are similar to those of central galaxies in non-fossil systems \citep{LaBarbera2009}. Moreover, the absence of large gradients in the metallicity radial profiles rules out the hypothesis of the monolithic collapse for BGGs in FGs \citep{Eigenthaler2013}. In summary, these observational properties indicate that BGGs in fossil and non-fossil systems show similar properties. The only relevant difference is the fraction of light enclosed in the BGG. This shows that BGGs in fossil groups may have formed via similar (but perhaps more efficient) physical mechanisms to non-fossil ones. 

All these observational properties can be explained in terms of the formation scenario of FGs. Numerical simulations show that the halo of a FG comprises half of its mass at z $>$ 1 \citep{DOnghia2004,DOnghia2005,Dariush2010}. Then, it grows via minor mergers alone, accreting only one third of the galaxies of regular groups or clusters \citep{vonBenda-Beckmann2008}. Moreover, in simulations, FGs always show an assembled mass that is, on average, higher than non-fossil systems at any redshift \citep{Dariush2007}.  These simulations seem to favor what is considered to be the ``classical'' formation scenario for FGs. They are thought to be very old systems that were able to assemble the majority of their mass at high $z$, where the $M^{\ast}$ galaxy population has been cannibalized by the BGG. 

Nevertheless, the formation of the BGG could be a long-term process. \citet{Diaz-Gimenez2008} claim that the last major merger for the BGG occurs at a later time in fossil than in non-fossil systems. \citet{Gozaliasl2014} suggest that the BGGs of fossil systems are the result of multiple mergers of $M^\ast$ galaxies in the past 5 Gyr. Moreover, \citet{vonBenda-Beckmann2008} claim that the fossil phase could be only transitional and that the interaction with other groups or clusters could erase the gap in magnitude. 
There is another possible scenario that is completely different from all of those mentioned above. This is the so-called {\it \emph{failed group}} scenario, which was first proposed by \citet{Mulchaey1999}. In this scenario the gap in magnitude is not due to the evolution of the system; rather, it appears by chance during the formation of the system itself. However, recent simulations matching subhalo abundance \citep{Hearin2013} seem to indicate that this scenario is not a good representation of reality.

The short formation time described in the classical scenario would give fossil systems enough time to merge all $M^{\ast}$ galaxies to form the massive central galaxy. The $M^{\ast}$ galaxies are the natural candidates for the merging process, since the dynamical friction---which is responsible for the decay of the orbits \citep{Chandrasekhar1943}---is higher for more massive satellites. Moreover, FGs are supposed to host $M^{\ast}$ galaxies in more radial orbits, and this can boost the efficiency of the merging process \citep{Sommer-Larsen2006,Boylan-Kolchin2008}. This means that fossil and non-fossil systems should have different luminosity functions (LFs).

The LF gives the number density of galaxies per luminosity interval, and it is a very powerful tool for studying the galaxy population in groups and clusters. A recent review of the principal results can be found in \citet{Johnston2011}.
In the case of FGs, because there are so few known systems, the majority of publications have analyzed the LFs of individual FGs. In particular, each analyzed system seems to show a peculiar LF that does not accord with the others \citep[see][]{Khosroshahi2006,MendesdeOliveira2006,Aguerri2011,Adami2012,Khosroshahi2014}. Thus, a systematic and homogeneous study is still required. For this reason, we present a large study here of a sample of 110 systems, containing 19 confirmed fossils. The criteria in the definition of fossils are those reported in \citet{Zarattini2014}. With this unique data set, we are able to present the first study of the dependence of the LF on the magnitude gap in group- and cluster-sized systems within half the $R_{200}$ radius.

This work is part of the FOssil Group Origins (FOGO) project. This is a multiwavelength study focused on the sample of 34 FG candidates proposed by \citet{Santos2007}. A detailed overview of the FOGO project is presented in the first paper of the series, \citet[][ hereafter FOGO I]{Aguerri2011}. In \citet[][ hereafter FOGO II]{Mendez-Abreu2012}, we explored the properties of the central galaxies; in \citet[][ hereafter FOGO III]{Girardi2014}, we presented the study of the L$_{\rm X}$-L$_{\rm opt}$ relations; and the characterization of the sample was given in \citet[][ hereafter FOGO IV]{Zarattini2014}. 
The structure of this paper is as follows. Section \ref{sample} is devoted to describing the sample; Sect. \ref{LF} shows how the LFs are calculated; Sect. \ref{results} describes the dependence of the LFs on the magnitude gap; and Sects. \ref{discussion} and \ref{conclusions} present the discussion and the  conclusions, respectively.

For this work, the adopted cosmology is $H_0 = 70$ km s$^{-1}$ Mpc$^{-1}$, $\Omega_{\Lambda} = 0.7$, and $\Omega_M = 0.3$.

%%%%%%%%%%%%%%%%%%%%%%%%%%%%%%%%%%
%%%%%%%%%%%%% SECTION 2%%% %%%%%%%%%%%%
%%%%%%%%%%%%%%%%%%%%%%%%%%%%%%%%%%
\section{Description of the sample}
\label{sample}

We used two samples of galaxy aggregations to analyze their LF and their dependence on the magnitude gap.
The first sample (hereafter S1) is composed of 34 groups and clusters of galaxies selected by \citet{Santos2007} and analyzed in detail in FOGO IV. These systems were selected as FG candidates from the Sloan Digital Sky Survey Data Release 5 \citep[SDSS DR5;][]{Adelman-McCarthy2007}, and they present a wide range in redshift ($0 < z < 0.5$). However, in our detailed analysis of this sample using deep $r$-band images and multi-object spectroscopy, it is shown that only 15/34 systems meet at least one of the definitions of fossil systems given by \citet{Jones2003} and \citet{Dariush2010}. The former authors claim that a cluster or group of galaxies is fossil if it has a gap in magnitude  larger than 2 in the $r$-band between the two brightest member galaxies ($\Delta m_{12} \ge 2$). The latter authors suggest that a system is fossil if the gap in the $r$-band between the first- and fourth-ranked galaxies is larger than 2.5 ($\Delta m_{14} \ge 2.5$). Both quantities are defined within half the virial radius.

The need for a second sample comes from the mean value of $\Delta m_{12}$ being $\sim$1.5 in the S1 sample. Only four systems have gaps lower than 0.5. We are interested in studying the dependence of the LF on the $\Delta m_{12}$ key parameter, so we need to extend the sample toward systems with small $\Delta m_{12}$.

For these reasons, we used a second sample (hereafter S2), taken by \citet{Aguerri2007}. These systems were selected as all the galaxy aggregations with known redshift at $z < 0.1$ from the catalogs of \citet{Abell1989}, \citet{Zwicky1961}, \citet{Boehringer2000}, and \citet{Voges1999} that were mapped in the SDSS DR4 \citep{Adelman-McCarthy2006}. This selection results in 88 systems. Of these, we used only those for which the two brightest members were spectroscopically confirmed, which limits the final number of systems from this sample to 76. For this sample, the mean value of $\Delta m_{12}$ is $\sim 0.7$, with a standard deviation of $\sim 0.5$. 

The general properties of the two samples are presented in \citet{Zarattini2014} for the S1 sample and in \citet{Aguerri2007} for the S2 sample. Both samples are mapped in the SDSS, and we used their $r$-band model magnitude \citep[see][]{Stoughton2002}.  The selection was done using SDSS DR5 for the S1 sample and DR4 for the S2 sample, but the magnitude used for this work were taken from the more recent SDSS-DR7 \citep{Abazajian2009}. These magnitudes were corrected for galactic extinction and K-correction. The former was obtained by using the $r$-band extinction parameter provided by SDSS. The latter was computed following the \citet{Chilingarian2010} and \citet{Chilingarian2012} prescriptions. Moreover, the $R_{200}$ radius of each system was computed using X-ray data from the ROSAT satellite (see FOGO III and FOGO IV). The LFs of this work were computed within half the obtained $R_{200}$ radius.

The X-ray luminosity of S1 ranges between $41.9 \le {\rm\log}\,(L_{\rm X, S1}\, [{\rm erg\, s^{-1}}]) \le 45.1$, whereas that of the S2 sample varies in the range $ 41.6 \le {\rm\log}\, (L_{\rm X,S2}\,{\rm [erg\, s^{-1}]}) \le 45.2$. The median values of the X-ray luminosity of the S1 and S2 samples are ${\rm\log}(L_{\rm X, S1}\,[{\rm erg\, s^{-1}}])=44.1\pm0.7$ and ${\rm\log}(L_{\rm X, S2}\,[{\rm erg\, s^{-1}})]=43.8\pm0.7$. Masses can be obtained using Eq. 3 of \citet{Rykoff2008}, after an adequate cosmology correction, and it varies in the ranges $13.1 \le {\rm\log}\,(M_{200,{\rm S1}}\,[{\rm M}_{\odot}]) \le 15.0$ and $12.9 \le {\rm\log}\, (M_{200,{\rm S2}}\,[{\rm M}_{\odot}]) \le 15.1$ for the S1 and S2 samples, respectively. The median values of the masses are ${\rm\log}\,(M_{200,{\rm S1}}\,[{\rm M}_{\odot}])=14.6\pm0.4$ and ${\rm\log}\,(M_{200,{\rm S2}}\,[{\rm M}_{\odot}])=14.4\pm0.4$, respectively. Finally, the velocity dispersion of the S1 and S2 galaxy systems span the ranges $250 \le \sigma_{v,{\rm S1}} \,[{\rm km\, s^{-1}}] \le 1200$ and $250 \le \sigma_{v,{\rm S2}} \,[{\rm km\, s^{-1}}] \le 1000$, respectively. The median values are $759\pm253$ km s$^{-1}$ and $557\pm170$ km s$^{-1}$, respectively. The Kolmogorov-Smirnov test was applied to the X-ray luminosities, masses, and velocity dispersion. Results from this test indicate that the S1 and S2 samples do not come from the same parent distribution. However, we are not directly comparing these two subsamples, since we want to compute the LFs in bins of $\Delta m_{12}$. In Sect. \ref{caveats} we show that the four analyzed subsamples actually come from the same parent distribution.

%median(log(Lxa))=43.8\pm0.7, min=41.6, max=45.2
%median(log(Lx_FG))=44.1\pm0.7, min=42.0, max=45.1

%median(logM200a))=14.4\pm0.4 , min=12.9, max=15.1
%median(log(M200_FG)=14.6\pm0.4, min=13.1, max=15.0

\subsection{Magnitude gap determination}
\label{gap}

For determining $\Delta m_{12}$ in the S1 sample we proceeded as follows.  We considered the four brightest galaxies of the systems within the spectroscopically confirmed members and possible members (see FOGO IV for both definitions), and then for each of these galaxies, we computed the magnitude as the mean value of three different magnitudes. We used the model and the Petrosian magnitude from the SDSS \citep{Stoughton2002} and the MAG-BEST magnitude by obtained analyzing our own deep $r$-band images with SExtractor \citep{Bertin1996}. We used this mean value to compute both $\Delta m_{12}$ and $\Delta m_{14}$ and used the standard deviation of the mean value for computing the uncertainties. The detailed procedure can be found in the FOGO IV paper.

For the S2 sample, we used the same methodology to obtain $\Delta m_{12}$ and $\Delta m_{14}$ except that we did not have our own photometric images. Thus, we used the mean value of the model and Petrosian magnitudes alone.

%%%%%%%%%%%%%%%%%%%%%%%%%%%%%%%%%%
%%%%%%%%%%%%% SECTION 3%%% %%%%%%%%%%%%
%%%%%%%%%%%%%%%%%%%%%%%%%%%%%%%%%%
\section{Galaxy luminosity function determination}
\label{LF}

\subsection{Luminosity functions of individual systems}
\label{ind_LF}

\begin{figure}
\includegraphics[width=0.5\textwidth]{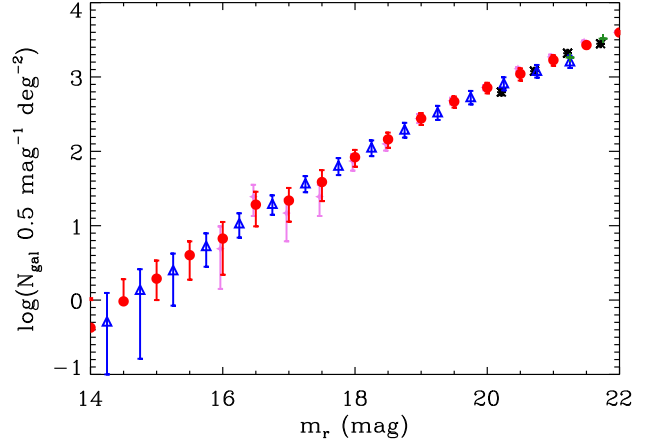}
\caption{Galaxy background estimations. Red filled circles represent the background used in this work, black asterisks are taken from \citet{Capak2004}, blue triangles from \citet{Yasuda2001},  violet stars from \citet{Huang2001}, and green plus signs from \citet{Metcalfe2001}.}
\label{BACK}
\end{figure}

There are two methods that are widely used in the literature to compute the  LFs of individual systems: the spectroscopic and the photometric ones. The former is, in principle, the most accurate. It is based on an extended redshift catalog, which allows for a detailed study of the system membership. Nevertheless, it requires a large amount of observational time and generally larger telescope apertures than the photometric one. This method has mainly been applied to the study of nearby individual clusters \citep[e.g., ][]{Rines2008,Agulli2014}. The photometric method requires less observing time and consists in computing the galaxy number counts, as a function of magnitude, within the system region and in a region of the sky in which no structure is present. The difference between these two quantities represents the galaxy system LF. It is a statistical method, so its main problem is that wherever the background field is located, it will not be exactly the same background of the system itself, mainly because of cosmic variance. Moreover, if the system is poor, it could have very limited contrast with respect to the background, leading to large uncertainties in the LF.

The LFs of individual systems, computed in half the $R_{200}$ radius, were obtained by using a hybrid method. This procedure uses both the photometric and spectroscopic information that we have for each system. The galaxy LF of each system in the j-th magnitude bin is given by
\begin{equation}
{\rm \phi_j}=N_{{\rm m},j}+(N_{j}-N_{{\rm v},j}) \times P_{j},
\label{eq_lf}
\end{equation} 
where $N_{{\rm m},j}$ is the number of spectroscopically confirmed members, $N_{j}$  the total number of galaxies, and $N_{{\rm v},j}$  the number of galaxies with recession velocity measurements. Finally, $P_{j}$ is given by

\begin{equation}
P_{j}=(N_{j}-N_{{\rm b},j})/N_{j},
\label{eq_back}
\end{equation}
where $N_{{\rm b},j}$ is the number of background galaxies. We notice that $P_j$ represents the probability that a galaxy would be considered in the LF when only photometry is available. The background was obtained by averaging four different fields used in the literature \citep{Yasuda2001,Metcalfe2001,Huang2001,Capak2004}. The resulting background distribution is shown in Fig.~\ref{BACK}. The completeness limit of our LFs is $r=21.5$ mag, and it is a conservative choice since the nominal completeness of SDSS DR7 $r$-band is 22.2 mag. 

The advantage of using this methodology is that we obtained a quasi-spectroscopic LF for the brightest bins, where the magnitude gap arises. In fact, the S1 sample is $\sim$85\% complete down to m$_r$=17 and the S2 sample is $\sim$90\% complete down to m$_r$=17.5 (see Fig. 4 of \citet{Zarattini2014} and Fig. 1 of \citet{Aguerri2007}). Then, when we move to fainter magnitudes, the number of measured redshifts decays rapidly and the LF is dominated by the statistical background subtraction. What we obtained is a hybrid LF that is more accurate at the bright end than the photometric one, but that is not as time consuming in terms of observations as a full spectroscopic LF in the faint end.
The uncertainty associated to the LF is calculated using the error propagation of the terms in Eq. \ref{eq_lf} and adding in quadrature the cosmic variance following \citet{huang1997}.

\begin{figure*}
\centering
\includegraphics[width=\textwidth]{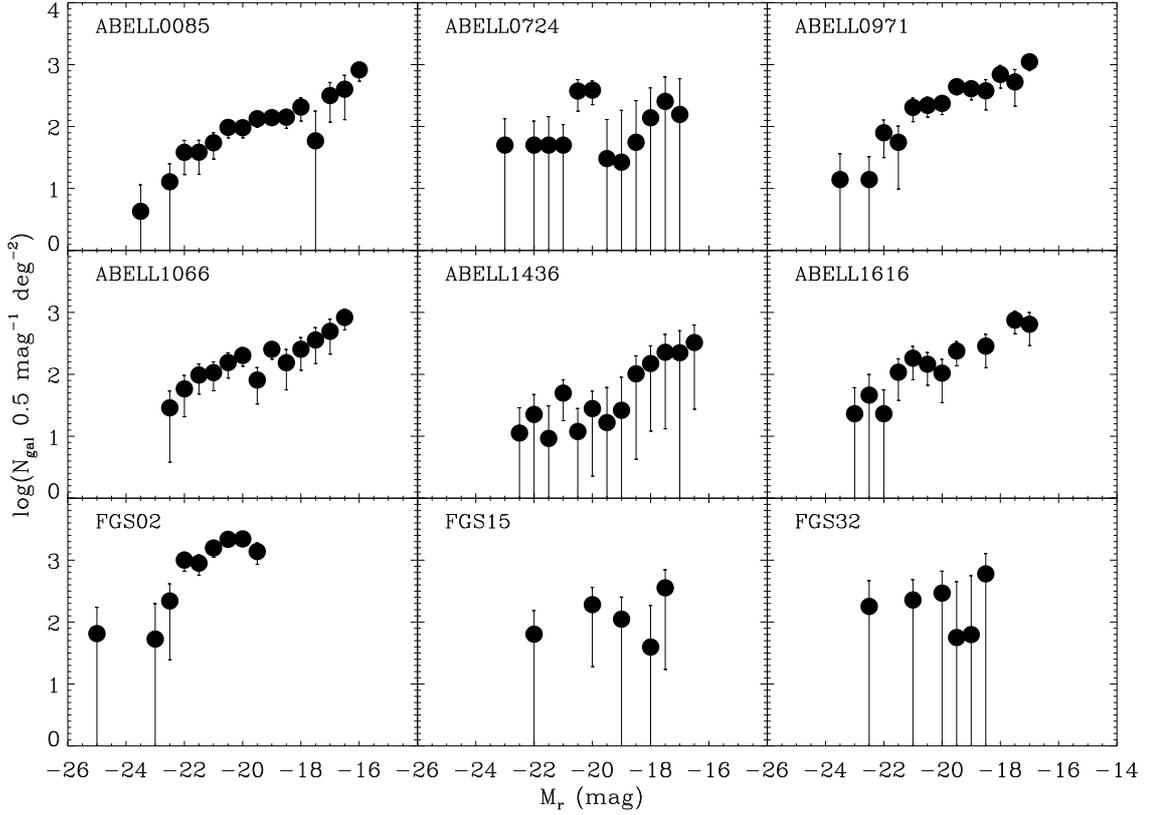} 
\caption{Examples of individual LFs for 9 systems taken from our sample. For the more massive ones, such as ABELL0085 and FGS02, the LFs are determined well, whereas for less massive systems, such as ABELL0724 and FGS15, the LFs have large uncertainties.}
\label{fig_ind_LF}
\end{figure*}

In Fig. \ref{fig_ind_LF} we show some examples of the individual LFs calculated with this procedure. For some rich objects, the individual LFs are clearly defined (e.g., Abell 1066 or FGS02), but for poorer systems the LFs present large uncertainties (e.g., Abell 724 or FGS15). All the LFs presented in this work were computed excluding the BGGs, as is usually proposed in the literature.

\subsection{Stacked luminosity functions}
\label{stack}

We stacked the individual LFs for all the S1 and S2 systems with $z\le0.25$ to deal with small numbers. This cut in redshift was needed to guarantee that all systems will reach at least a magnitude of $M_r = -19.5$ (given a completeness limit of the SDSS of $m_r=21.5$) so that they can be normalized using the region $-21 \le M_r \le -19.5$. As noted by other authors \citep[e.g.,][]{Popesso2005}, stacked LFs are not only a useful tool for checking the universality of the LF, but they are useful for calculating the LF of systems with high accuracy when the individual ones have poor statistics.

There are different methods in the literature of stacking LFs. We used the one proposed by \citet{Colless1989}, in which the stacked LF can be obtained by combining the individual ones according to the formula 

\begin{equation}
\phi_{cj}=\frac{N_{c0}}{m_j}\sum_i \frac{N_{ij}}{N_{i0}},
\label{eq_stacking}
\end{equation}
where $\phi_{cj}$ is the number of galaxies in the j-th bin of the stacked LF, $N_{ij}$  the number of galaxies in the j-th bin of the i-th individual system's LF, $N_{i0}$  the normalization of the i-th system LF in the region $-21 \le M_r \le -19.5$, $m_j$  the number of systems contributing to the j-th bin, and $N_{c0}$ is the sum of all the normalizations ($N_{c0}=\sum_i N_{i0}$).

The formal errors are computed according to

\begin{equation}
\delta \phi_{cj}=\frac{N_{c0}}{m_j}\left[\sum_i \left(\frac{\delta N_{ij}}{N_{i0}}\right)^2\right]^{1/2},
\end{equation}
where $\delta \phi_{cj}$ and $\delta N_{ij}$ are the formal errors in the LF's j-th bin for the composite and i-th system, respectively.

The Colless method can be safely used under a few conditions: first of all, the magnitude limit of all stacked systems must be fainter than the values used for normalization ($M_r=-19.5$ in our case). Second, the normalization region should be large enough to be representative of the richness of the systems. Finally, the number of systems contributing to each bin should be statistically relevant. 

\begin{figure}
\includegraphics[width=0.5\textwidth]{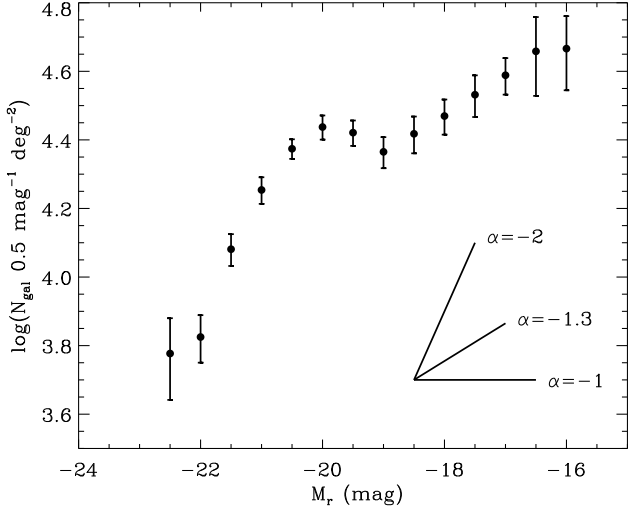}
\caption{Stacked LF for all the systems of our sample with $z \le 0.25$. The solid lines represent the faint-end slope for $\alpha=-2.0$, which is the value predicted by standard CDM theories, $\alpha= -1.3$, obtained from our fit, and $\alpha=-1.0$, which is the value for a flat LF.}
\label{LF_whole}
\end{figure}

As introduced at the beginning of this section, we applied this method to all the systems in our sample with $z \le 0.25$. Using this subsample, we satisfied all the requirements for applying the Colless method. The total number of systems turns out to be 102, and the resulting composite LF is presented in Fig. \ref{LF_whole}.  We fit neither a single nor a double Schechter function to the data, because the Spearman test told us that none of them was representative of the data. We fit an exponential function to the faint end of the LFs using the last five points. The form of the exponential is $10^{km}$, where $m$ represents the magnitude and $k$ is related to Schechter's $\alpha$ parameter by 
\begin{equation}
\alpha=-\left(\frac{k}{0.4}+1\right).
\end{equation}
The resulting faint-end slope is $\alpha=-1.27 \pm 0.11$. Hereafter, all the presented exponential slopes were fitted using the five faintest points of each LF.

%%%%%%%%%%%%%%%%%%%%%%%%%%%%%%%%%%
%%%%%%%%%%%%% SECTION 4%%% %%%%%%%%%%%%
%%%%%%%%%%%%%%%%%%%%%%%%%%%%%%%%%%
\section{Dependence of the luminosity function on the magnitude gap}
\label{results}

\begin{table}
\tabcolsep=0.5cm
\caption{Best-fitting parameters of a Schechter fit to the regular (top) and relative (bottom) LFs. Reported uncertainties represent the 99\% confidence level (c.l.) of each parameter.}
\label{params}
\begin{tabularx}{0.46\textwidth}{ccc}
\tiny
\\
\hline
\hline
$\Delta m_{12}$ & $M^{\ast}$& $\alpha$ \\ 
\hline
\vspace{0.2cm}
          $\Delta m_{12} < 0.5$  & $-22.30^{+0.61}_{-0.70}$& $-1.23^{+0.09}_{-0.10}$     \\ 
          \vspace{0.2cm}
0.5 $\le \Delta m_{12} < 1.0$  & $-22.16^{+1.06}_{-0.83} $& $-1.13^{+0.12}_{-0.11} $        \\ 
\vspace{0.2cm}
1.0 $\le \Delta m_{12} < 1.5$ & $-21.40^{+1.19}_{-1.53}$& $-0.90^{+0.52}_{-0.22}$\\
$\Delta m_{12} \ge 1.5$      & $-21.04^{+0.43}_{-0.52}$& $-0.78^{+0.26}_{-0.15}$         \\ 
\tiny
\\
\hline
\hline
            $\Delta m_{12}$ & $M^{\ast}$& $\alpha$ \\
            \hline
            \vspace{0.2cm}
$\Delta m_{12}< 0.5$  & $0.05^{+0.86}_{-1.4} $& $-1.26 ^{+0.10}_{-0.10}$\\ 
\vspace{0.2cm}
0.5  $\le \Delta m_{12} < 1.0$  & $1.59^{+0.53}_{-0.68}$        & $-1.03^{+0.13}_{-0.10} $\\ 
\vspace{0.2cm}
1.0 $\le \Delta m_{12} < 1.5$ & $1.95^{+1.04}_{-1.50}$&$-0.93^{+0.28}_{-0.17} $ \\
\vspace{0.2cm}
$\Delta m_{12}\ge 1.5$ & $2.85^{+0.55}_{-0.64}$  & $-0.77^{+0.32}_{-0.15}$       \\ 

\end{tabularx}
\end{table}

\begin{figure}
\includegraphics[width=0.5\textwidth]{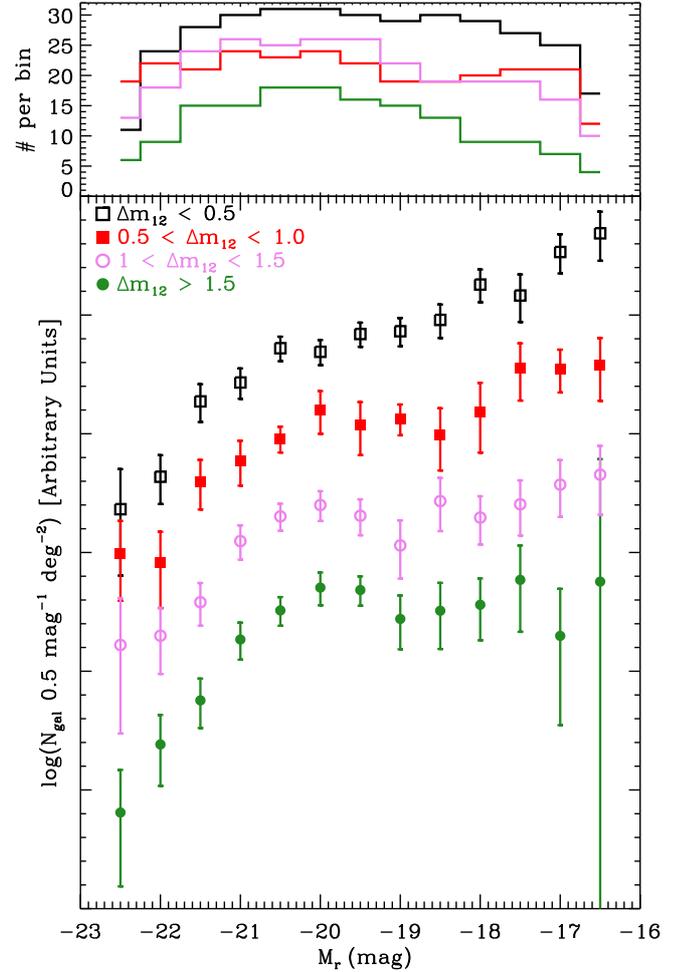}
\caption{Lower panel: Stacked LFs of systems with different gaps in magnitude. Black open squares are systems with gap $\le$ 0.5, red filled squares represent systems with $0.5 < \Delta m_{12} \le 1.0$, violet open circles indicate systems with $1 < \Delta m_{12} \le 1.5$, and green filled circles are systems with gap $\ge$ 1.5. The four LFs have been moved by an arbitrary offset for display purposes. Upper panel: histogram of the number of systems that are contributing to each bin. The color code is the same as in the lower panel.}
\label{LF_gap1}
\end{figure}

We divided the sample of the 102 systems with $z \le 0.25$ into four subsamples, which differ from one another in the value of $\Delta m_{12}$. The first subsample is composed of 31 systems with $\Delta m_{12} < 0.5$, the second of 24 systems with $0.5 \le \Delta m_{12} < 1$, the third of 26 systems with $1 \le \Delta m_{12} < 1.5,$ and the fourth of 21 systems with $\Delta m_{12} \ge 1.5$. This division is arbitrary and was done in order to have a statistically significant number of systems in each range of $\Delta m_{12}$ and to trace the dependence of the LF with $\Delta m_{12}$ in the best possible way. 
In Fig. \ref{LF_gap1} we show the stacked LFs for the four subsamples. Qualitatively, the slope of the bright end is similar for the four LFs, but not the faint-end one. In particular, the systems with $\Delta m_{12} < 0.5$ show a steeper faint end than those with $\Delta m_{12} \ge 1.5$, whereas the other two subsamples represent intermediate cases. To quantify this effect, we fit a single Schechter function to each LF shown in Fig. \ref{LF_gap1}. In this case, the Spearman test confirms that a single Schechter function is a reasonable representation of the data of each subsample. 

\begin{figure}
\includegraphics[width=0.5\textwidth]{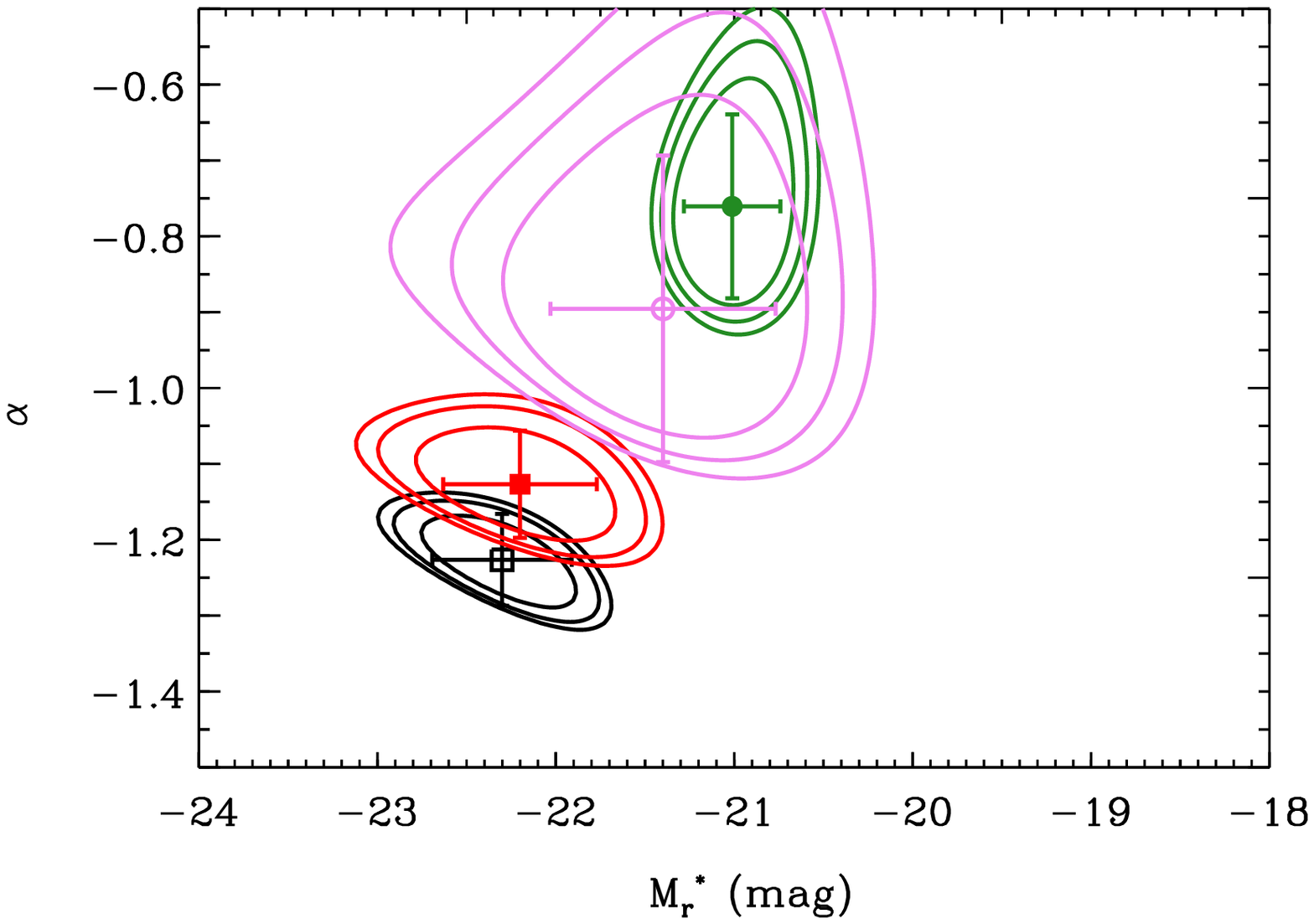}
\caption{Uncertainty contours for the Schechter fits of Fig. \ref{LF_gap1}. Contours represent 68\%, 95\%, and 99\% c.l. and the color and symbol codes are the same as in Fig. \ref{LF_gap1}. The error bars are the 1-$\sigma$ uncertainties of the Schechter fit as reported in Table \ref{params}.}
\label{mstar_chi}
\end{figure}

The Schechter function \citep{Schechter1976} is the most accepted expression to describe the galaxy LF parametrically. Its formulation can be written as follows:

\begin{equation}
\phi(M) dM= \phi^\ast (10^{0.4(M^\ast-M)}){^{(\alpha+1)}} {\rm exp}(-10^{0.4(M^\ast-M)})dM,
\end{equation}
where $\phi^\ast$ is a normalization factor defining the overall density of galaxies and $M^\ast$ is the characteristic magnitude. The parameter $\alpha$ describes the faint-end slope of the LF, and it is typically negative. 
The results of the fit are shown in the upper part of Table \ref{params}. There are differences in both the bright and the faint ends. In the former, the larger the gap, the fainter the $M^{\ast}$. In the latter, the larger the gap, the flatter the $\alpha$. We plotted the 68\%, 95\%, and 99\% confidence level (c.l.) contours for $M^\ast$ and $\alpha$ of the Schechter fit in Fig. \ref{mstar_chi}. We refer to LFs computed using this method as ``regular'' LFs.

Moreover, for the faint end, we fit an exponential function as well. This check is useful for two reasons: it helps to quantify the effect of the known degeneracy between $M^\ast$ and $\alpha$, and it can be useful to compare the results with other LFs that are not well described by a Schechter function. The obtained values for the exponential faint-end slope are $\alpha=-1.43 \pm$ 0.12, $-1.39 \pm 0.17$, $-1.18 \pm 0.31$, and $-1.02 \pm 0.32$ for the four LFs with an increasing gap, respectively. These values are higher than for Schechter's $\alpha$ parameter, but the trend is the same. However, the differences in $\alpha$ between the four slopes are not statistically significant because of the large uncertainties of the stacked galaxy LFs for systems with larger $\Delta m_{12}$. 

The BGGs in our sample span a four-magnitude range \citep[see][]{Zarattini2014}. This can affect the shape of the bright end of the stacked LFs, because the individual LFs are not aligned in magnitude. To avoid any effect associated with the stacking of different galaxy populations, we have computed the stacked LFs as a function of the relative magnitude. This was obtained by calculating the differences between the magnitude of the galaxies and the BGG ($\Delta M_r=M_r - M_{r,BGG}$)
for each system. Using this method, all the BGGs are located at $\Delta M_r=0$.  We refer to the resulting LF as the ``relative'' LF. In this picture the value of $M^\ast$ loses its physical meaning, but we are interested in highlighting the differences in the Schechter parameters between the four subsamples, not in their absolute values. 

There are two differences between the methodology that we applied for the regular LF and for the relative LF. The first difference is the already mentioned shift of the magnitudes in the calculation of relative LFs. The second difference is that, for the relative LF, spectroscopically confirmed zero-galaxy bins enter into the stacking formula (Eq. \ref{eq_stacking}) for each bin of magnitude and for each system. In contrast, for the regular stacking, this was not possible because of the large difference in the magnitude of the central galaxies. In fact, the stacking is actually a computation of the mean value of the LF in each bin, but when the dominant parameter is the absolute magnitude, mixing massive systems with small groups could be misleading. If a group is dominated by a central galaxy of M$_r=-22$ and has, for example, a spectroscopically confirmed two-magnitude gap, when we try to stack it with a massive cluster whose central galaxy has M$_r=-25$, the spectroscopically confirmed gap of the group would affect a part of the cluster that is three to five magnitudes fainter than the central galaxy of the cluster itself. This part, assuming a Schechter profile for the distribution of galaxies, would probably be located beyond the elbow ($M^\ast$) of the LF of the cluster. Thus, we expect that in this region the cluster presents a large number of objects. Doing the stacking in this case would imply reducing the galaxies that are present in that bin by a factor of 2. Clearly, if the number of systems is more than two, as in our case, the effect would be softened, but we expect a flattening of the elbow region if we do not take this aspect into account.

\begin{figure}
\includegraphics[width=0.5\textwidth]{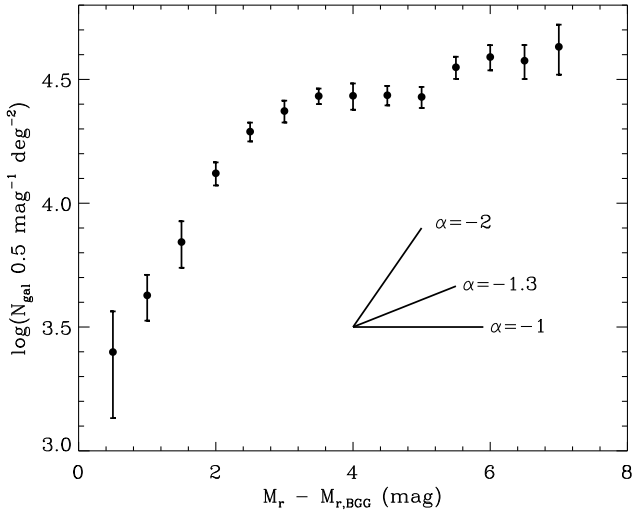}
\caption{Stacked relative LF for all the systems of our sample with $z \le 0.25$. The two solid lines represent the same $\alpha$ values as in Fig. \ref{LF_whole}.}
\label{LF_whole_shift}
\end{figure}

\begin{figure}
\includegraphics[width=0.5\textwidth]{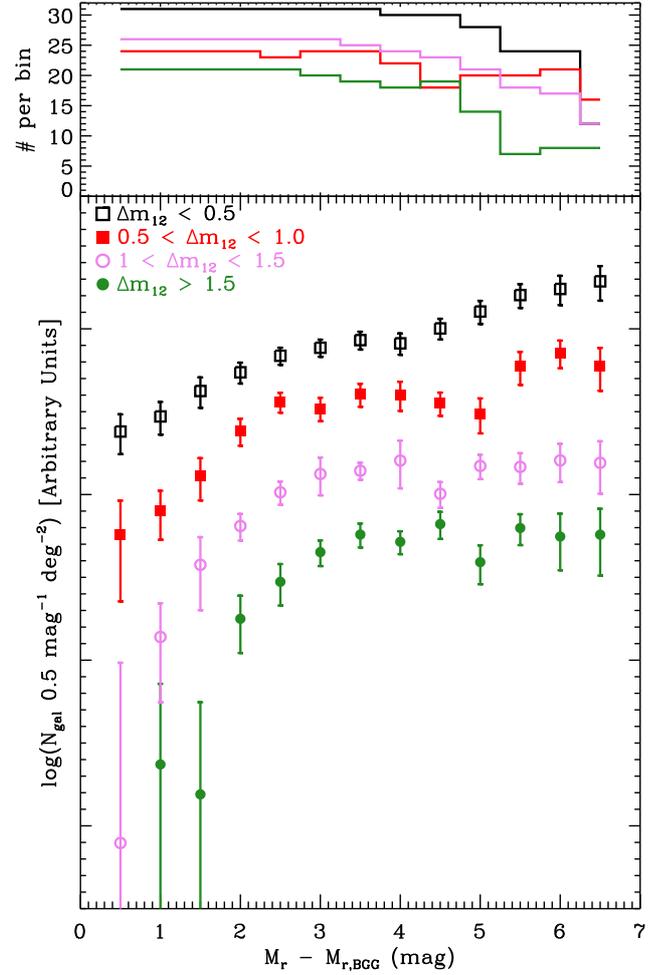}
\caption{Same as Fig. \ref{LF_gap1} but with relative LFs.}
\label{LF_gap_shift}
\end{figure}

\begin{figure}
\includegraphics[width=0.5\textwidth]{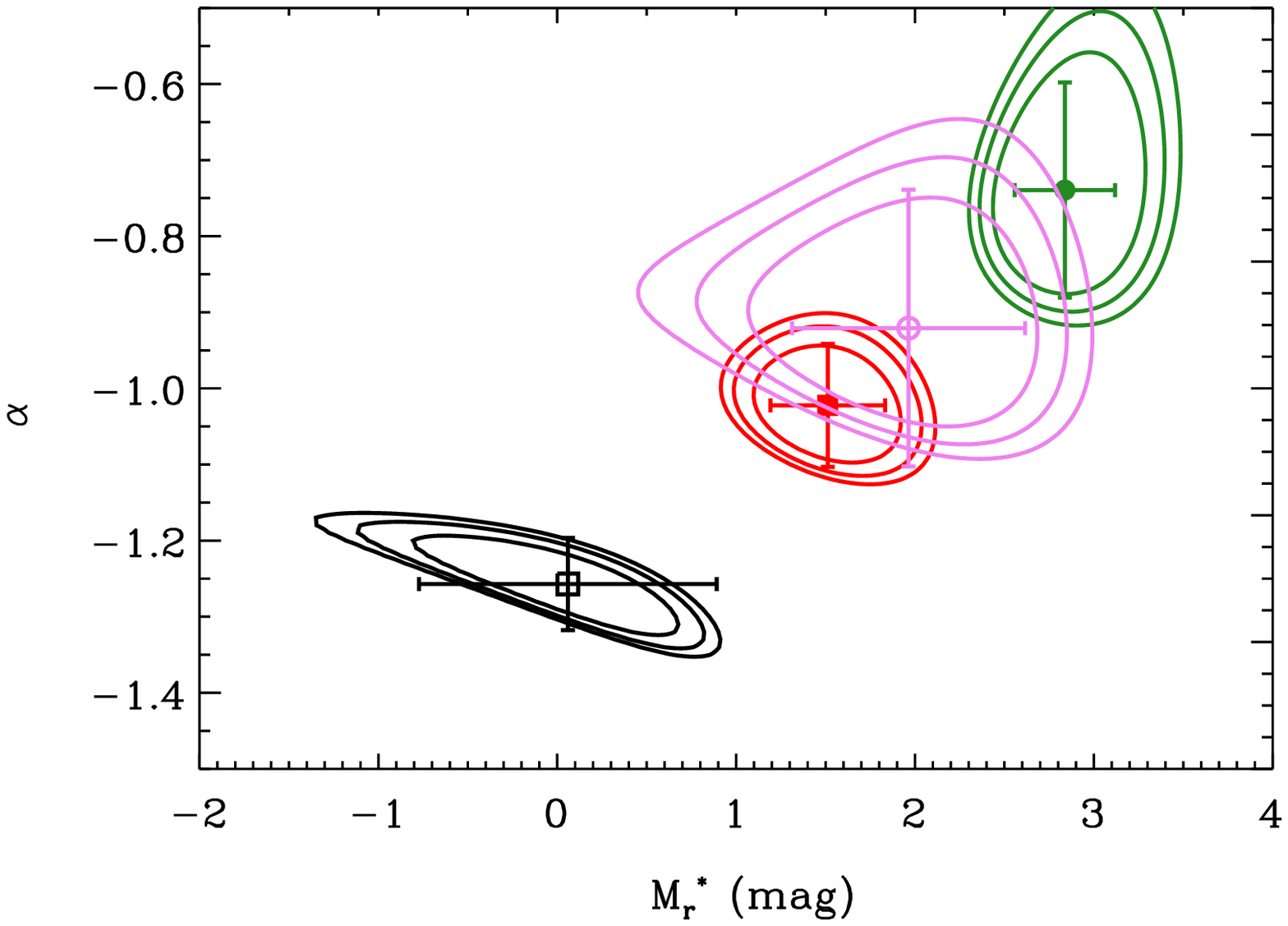}
\caption{Uncertainty contours for the Schechter fits of Fig. \ref{LF_gap_shift}. Contours represent 68\%, 95\%, and 99\% c.l. and the color and symbol codes are the same as in Fig. \ref{LF_gap1}. The error bars are the 1-$\sigma$ uncertainties of the Schechter fit as reported in Table \ref{params}.}
\label{mstar_chi_shift}
\end{figure}

In Fig. \ref{LF_whole_shift} we show the stacked relative LF for the whole sample of 102 systems with $z \le 0.25$. As we did for the regular LF, we fit an exponential to the faint end of the relative one. The resulting slope is $\alpha=-1.25 \pm 0.09$, which is compatible (within the uncertainties) with the value measured for the regular LF.
Figure \ref{LF_gap_shift} shows the relative LFs of the four subsamples with different magnitude gaps. Qualitatively, we found differences in both the bright and the faint ends of the four stacked relative LFs. We fit a single Schechter function to these relative LFs and show the obtained $M^\ast$ and $\alpha$ parameters in Table \ref{params}. Their uncertainties are shown in Fig. \ref{mstar_chi_shift}. The Schechter parameters of the smallest and largest magnitude gap regimes have a greater difference in value for the relative LFs than the regular LFs. Once again, we fit an exponential to the faint end of the four LFs. The resulting faint-end slopes are $-1.37 \pm 0.12$, $-1.43 \pm 0.13$, $-1.24 \pm  0.22$, and $-0.95 \pm 0.17$, moving from the smaller to the larger gap. In this case, the exponential fits also show differences at the 1-$\sigma$ level between the systems with $\Delta m_{12} < 0.5$ and $\Delta m_{12} \ge 1.5$. The larger uncertainties are due to the small number of data points available for the exponential fit. This fit has been performed as a test to break the degeneracy between M$^\ast$ and $\alpha$ that is obtained when both the bright and faint ends are fitted at the same time. The exponential fit of the LF faint end is statistically less significant. But, it has to be considered as a double check of the slope derived by fitting the full LF with a Schechter function.

In Fig. \ref{LF_gap_shift}, there are some points in the LFs that are located at a magnitude difference from the BGG that is smaller than the magnitude gap used to define the four subsamples. For example, in the sample with $\Delta m_{12} \ge 1.5$ the LF starts at $\Delta m_{12}$=1. This apparent contradiction is due to the definition of the magnitude gap. It is obtained by only using possible members. The selection of possible members cannot be made using a statistical background subtraction, because we need individual galaxy information. For this reason, in the determination of the gap we used a generous cut in photometric redshift to account for possible cluster members \citep[see][for details]{Zarattini2014}. Nevertheless, no photometric-redshift cut was applied for the computation of the galaxy LF. The methods are not incoherent, since the error bars of these peculiar points are always compatible with zero. These points should be considered as statistical fluctuations in the very bright part, but they do not affect the results owing to their large uncertainties.

\subsection{Dwarf-to-giant galaxy ratio}
It is important to remember that the stacked LFs is arbitrarily normalized, which means that looking at the absolute number of galaxies at both the bright and the faint ends of the LFs can be misleading. For this reason, we analyzed the so-called dwarf-to-giant galaxy ratio (DGR) for the four subsamples. We defined as giant galaxies those in the range $-22.5 \le M_r \le -20$, and as dwarf galaxies those in the range $-19 \le M_r \le -17$. The resulting DGR for the four subsamples with increasing magnitude gaps are $2.27 \pm 0.03$, $1.70 \pm 0.02$, $1.37 \pm 0.02$, and $1.33 \pm 0.02$.
%In the case of the relative LFs we defined as giant those galaxies with $M_r -M_{r,BGG} \le 3$ and as dwarf those galaxies with $4 \le M_r -M_{r,BGG} \le 6.5$. We find a DGR of 2.83$\pm$0.03, 2.47$\pm$0.02, 2.47$\pm$0.03 and 3.46$\pm$0.04 for increasing magnitude gap subsamples, respectively.
The DGR is very difficult to compare with the literature. For example, \citet{Popesso2005} defined galaxies in the range $-18 \le M_r \le -16.5$ as dwarfs and those with $M_r \le -20$ as giants, \citet{Sanchez-Janssen2008} considered as dwarfs those with $M_r > M_r^\ast +1$ and as giants those with $M_r < M_r^\ast$, and \citet{Weinmann2011} defined as dwarf galaxies those with $-16.7 > M_r > -19$ and as giant those galaxies with $M_r < -19$. While the exact value of the DGR is therefore highly arbitrary, the important result of this exercise is to show that we recover the trend previously found using the fits to the LF: the relative number of dwarfs systematically decreases in systems with progressively larger magnitude gaps.

%%%%%%%%%%%%%%%%%%%%%%%%%%%%%%%%%%
%%%%%%%%%%%%% SECTION 5%%% %%%%%%%%%%%%
%%%%%%%%%%%%%%%%%%%%%%%%%%%%%%%%%%
\section{Discussion}
\label{discussion}

\begin{figure}
\includegraphics[width=0.5\textwidth]{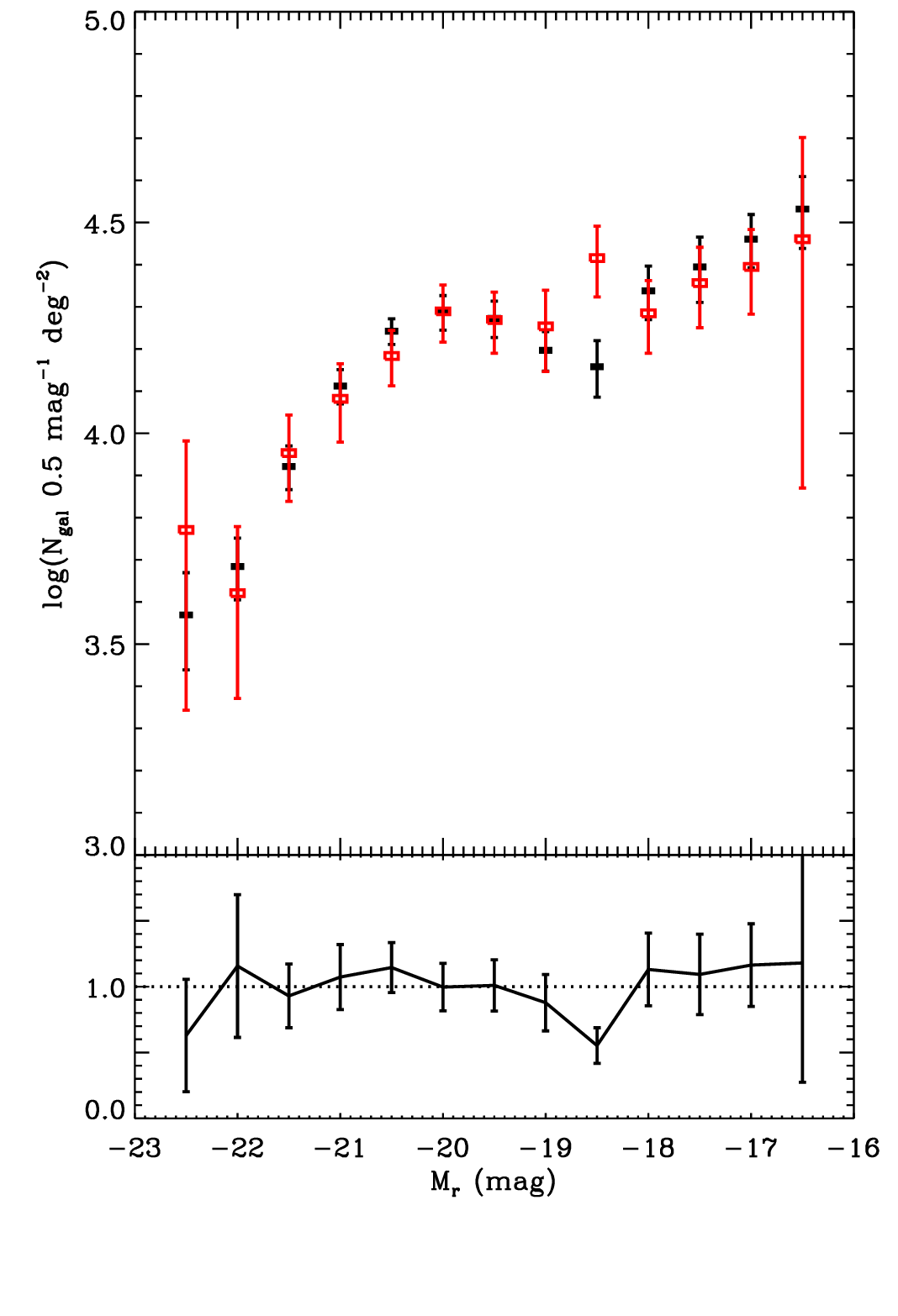}
\caption{Stacked LFs for both clusters (black filled rectangles) and groups (red open rectangles) in the upper panel. Here we considered objects with $\sigma _{\rm v} \le 580$ km s$^{-1}$ as groups and objects with $\sigma _{\rm v} > 580$ km s$^{-1}$ as clusters.The ratio of the two LFs is plotted in the lower one.}
\label{clus_vs_groups}
\end{figure}

\subsection{Caveats of the results}
\label{caveats}
In Figs. \ref{LF_gap1} and \ref{LF_gap_shift} we have stacked all the available systems, mixing clusters, and groups. This can affect the result, since in some of the four magnitude-gap bins we could be dominated by massive clusters, while in others the dominant systems could be groups. Differences can be found in the literature between the Schechter parameters of clusters and groups \citep[e.g., ][]{Zandivarez2011}. To test this aspect, we ran a Kolmogorov-Smirnov test, which confirmed that the distributions of $\sigma _{\rm v}$ (which is a mass proxy, as suggested by \citet{Munari2013}, and it has been obtained from L$_{\rm{X}}$) for the four subsamples that are not different. Moreover, we computed the median $\sigma _{\rm v}$ for the four subsamples. The resulting values are $\sigma _{\rm v}=557$, 591, 587, and 545 km s$^{-1}$, with standard deviations of 171, 159, 206, and 200 km s$^{-1}$, for the four subsamples ordered with increasing magnitude gap. These two tests indicate that the four subsamples show the same mean velocity dispersion, hence the same mean mass. We also divided the whole sample (S1+S2) into two mass bins, defining as groups those systems with $\sigma_{{\rm v}} \le 580$ km s$^{-1}$ and as clusters those with $\sigma_{{\rm v}} > 580 $ km s$^{-1}$. This value represents the median value of $\sigma_{{\rm v}}$ calculated over the whole S1+S2 sample. In Fig. \ref{clus_vs_groups} we present the stacked LFs according to the mass of the systems and the ratio between the two LFs. It can be seen that the LFs are in good agreement with one another, because the differences are always compatible with zero except for one point, M$_r = -18.5$. In fact, in this point there is a dip in the more massive systems, as already found by other authors \citep[see Fig. 11 in][]{Trentham2002}. In conclusion, the observed differences in the LFs do not seem to be related to the mass distribution of the systems in the four subsamples.

\begin{figure}
\includegraphics[width=0.5\textwidth]{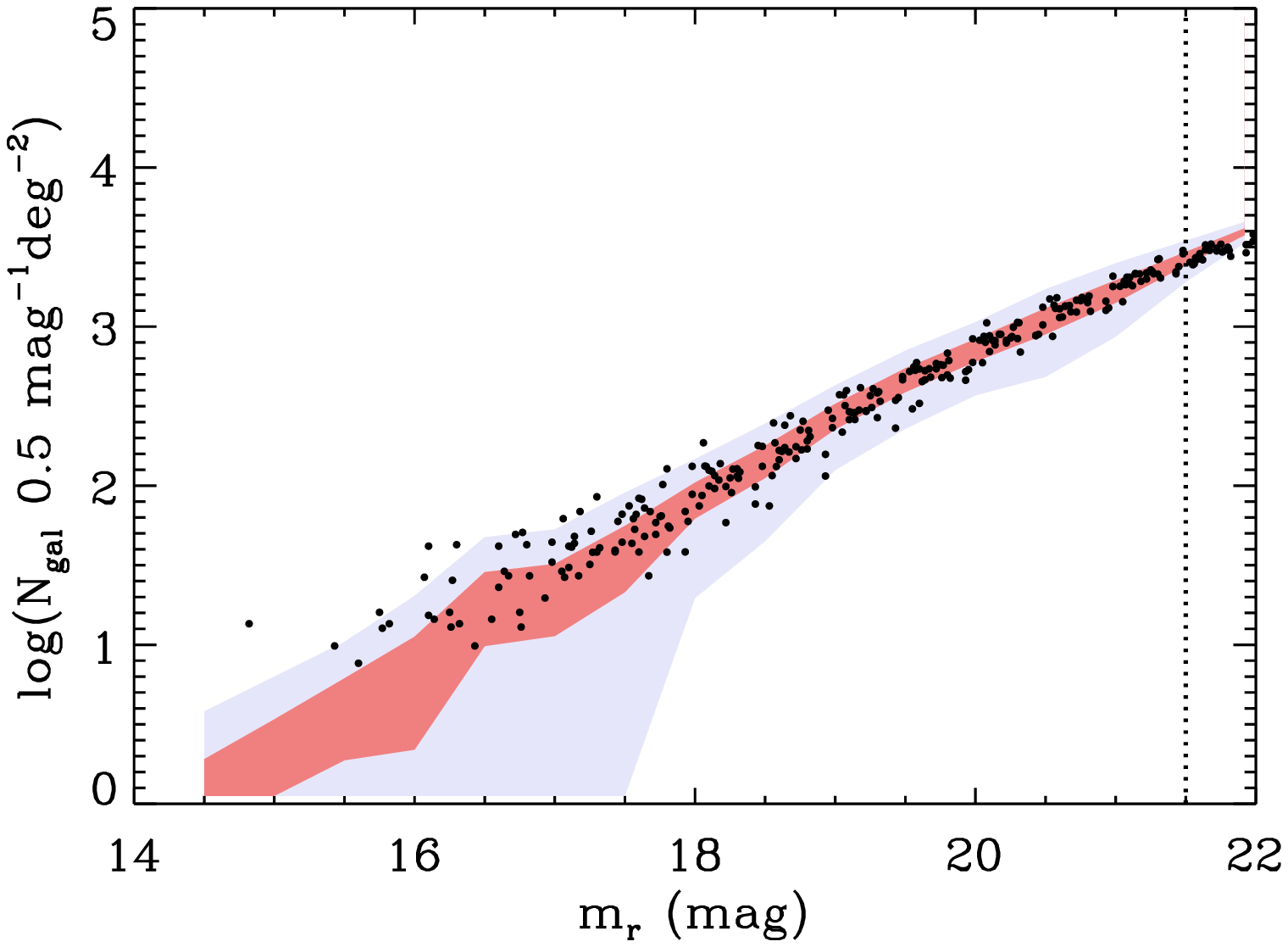}
\caption{Local background (black dots) superimposed on the 1-$\sigma$ and 3-$\sigma$ contours of the global background (red and gray shaded areas, respectively). The vertical dotted line is the completeness limit of this work.}
\label{test_BG}
\end{figure}

As pointed out in Sect. \ref{ind_LF}, we used a galaxy background in order to compute the galaxy LFs. We tested how the results were affected by changing the adopted galaxy background. We used the 26 systems of the S1 sample that are closer than $z=0.25$. These systems are representative of the whole S1+S2 sample in terms of mass ($\sigma _{\rm v}$), and their $R_{200}$ are smaller than 15 arcmin. We then calculated a local background for each system in an annulus between two to four times their $R_{200}$ radius. We divided the annulus into 20 regions of the same area, as proposed by \citet{Popesso2005}. Then, we counted the number of galaxies for each bin of magnitude in each sector, and finally we calculated the mean value of the local background by averaging all the sectors, using a sigma-clipping algorithm to exclude sectors that are at more than 3-$\sigma$ from the global mean value. In Fig. \ref{test_BG} we show the values of the local background for each system and the 1-$\sigma$ and 3-$\sigma$ uncertainties of the global background presented in Fig.~\ref{BACK}. It can be seen that up to m$_r=17$ the local and global backgrounds seem to disagree with one another, but this does not present a problem in our case because, as we already mentioned, we are using a hybrid method for the LF, and both the S1 and the S2 samples have a spectroscopic completeness of more than 85\% up to m$_r=17$. Moreover, for magnitudes brighter than $m_r=17,$ the local background method is less reliable, since bright galaxies are scarce, and a large area is needed to properly take them into account. For magnitudes between $m_r=17$ and $m_r=21.5$, which is our conservative completeness limit, the global and local backgrounds are in good agreement, so no large differences are expected in the calculation of the LF by varying the  background.

The stacking procedure computes a mean LF, using for each bin of magnitude only those clusters where the magnitude limit is fainter than each specific bin. Thus, each bin of the composite LF is formed by a different number of averaged points. At the faint end, the systems that are contributing are the closest in terms of redshift, since our sample is limited in apparent magnitude. However, the faint-end slope is not expected to change in the redshift range $0 \le z \le 0.25$ \citep{Gozaliasl2014}, thus we think that the method does not introduce any bias into the dependence of the faint-end slope on the magnitude gap. In the bright end, differences could arise due to the use of our hybrid method. Nevertheless, no differences are expected in the bright end when applying a fully photometric or a fully spectroscopic method, although the latter has smaller uncertainties. Thus, the use of a hybrid method should not affect the computation of the M$^\ast$ values, but it should help in reducing their uncertainties. We have plotted in Figs. \ref{LF_gap1} and \ref{LF_gap_shift} the histograms showing the number of systems per bin of magnitude used in computing the stacked LFs.

We also analyzed the differences in the redshift distributions of the four subsamples. The Kolmogorov-Smirnov test indicates that the four redshift distributions differ from one another. The median redshift value for subsamples with increasing magnitude gap are $z=$ 0.064, 0.077, 0.088, and 0.11, respectively. Nevertheless, the differences in redshift are small, and \citet{Gozaliasl2014} show that no evolution is expected in the faint-end slope for both fossil and non-fossil systems since $z=1$. Moreover, the lookback time at $z=0.1$ is $\sim$ 1 Gyr, which is a small amount of time to see an evolution. Thus, we think that the differences in redshift in the four subsamples do not represent a bias for our results.

To conclude our analysis of the possible caveats, we investigated how the uncertainties in the magnitude determination depend on the magnitude itself, and how this can affect the computation of the LFs. We used SDSS model magnitudes, so to constrain the uncertainties we analyzed the distribution of the ${\rm modelMagErr_r}$ parameter. The median uncertainty at m$_r=21.5$ -- our conservative completeness limit -- is 0.15 mag. Since our bins are 0.5 mag wide, we expect that the photometric uncertainties do not affect the results.

\subsection{Comparison with the literature}
We can use the regular LFs to compare the results of this work with other results in the literature for the $r$-band. For example, \citet{Popesso2005} found a slope of $\alpha=-1.30 \pm 0.06$ for the bright part ($M_r \le -18$) of the stacked LF of $\sim$100 clusters inside 1 Mpc, and a slope of $\alpha=-1.29 \pm 0.09$ for the same sample within 0.5 Mpc. In contrast, \citet{deFilippis2011} find a faint-end slope of $\alpha=-0.99^{0.01}_{0.02}$ for a stacked LF of $\sim$1500 systems. We can also try to compare our stacked LF with individual systems analyzed in the literature. For example,  \citet{Rines2008} analyzed spectroscopic LFs of Abell 2199 and the Virgo cluster, finding $\alpha=-1.02 \pm 0.05$ for the former and $\alpha=-1.28 \pm 0.06$ for the latter. These results, obtained with different techniques, are compatible with our faint-end slope of $\alpha= -1.27 \pm 0.11,$ with the only exception being \citet{deFilippis2011}. Finally, \citet{Barrena2012} find no difference in the $\alpha$ parameters in relaxed and unrelaxed clusters with $\alpha=-0.86 \pm 0.27$ and $\alpha= -0.99 \pm 0.21$. This result is of particular interest since fossil systems are thought to be old and, thus, more relaxed than non-fossils. 

The evolutionary state of the system could be seen as an explanation of our results, too. In fact, \citet{Iglesias-Paramo2003} studied two clusters (Coma and Abell 1367) with similar X-ray luminosities and redshifts and found significant differences in their faint-end slopes. The authors suggest that these differences could be explained by differences in the evolutionary state of the two clusters. Thus, if applying this result to our case, a possible explanation for the observed differences is that the $\Delta m_{12}$ parameter could be an indicator of the evolutionary state of a system. Either way, extended work on the substructure in fossil systems remains to be done, so differences in their evolutionary state cannot be proved.

It is also interesting to compare the result of the $\Delta m_{12} \ge 1.5$ subsample with the results found for fossil systems in the literature. In fact, in this subsample, 13 out of 21 systems are spectroscopically confirmed fossils. \citet{Khosroshahi2006} find $\alpha= -0.61 \pm 0.20$ for the background-corrected photometric LF within 0.5 $R_{200}$ and \citet{MendesdeOliveira2006} find  $\alpha= -0.64 \pm 0.30$ in $\sim$0.3 $R_{200}$, whereas in FOGO I we found $\alpha= -0.54 \pm 0.18$ for LFs within 1 Mpc. It can be seen that, despite the large uncertainties, all the results seem to point to a value of the faint-end slope that is higher than -1 for fossils. In this sense, our result agrees with the literature, since our faint-end slope in 0.5 $R_{200}$ for systems with large magnitude gaps is $\alpha= -0.78 \pm 0.12$ for the regular LF and $\alpha= -0.77 \pm 0.14$ for the relative LF. That our result is the lowest one can be interpreted as an effect of non-fossil systems being in this subsample, owing to its definition. The effect of this contamination would be a steepening of the LF. This agrees with our general result that the faint-end slope increases with the gap in magnitude. In a recent work, \citet{Lieder2013} analyzed the FG NGC 6482. They present a deep spectroscopic LF, down to $M_r=-10.5$. The faint-end slope they found was $\alpha=-1.32 \pm 0.05$ in one virial radius, which is steeper than the other works in the literature. However, their result is not directly comparable to the others, since they fit the faint-end slope down to a much deeper magnitude. Finally, in a very recent work, \citet{Wen2015} have analyzed the dependence of the bright end of the LF on the cluster dynamical state. They used the method presented in \citet{Wen2013} to create three subsamples of clusters with different dynamical states. They conclude that more relaxed clusters have fainter M$^\ast$. Thus, our results for the bright end would indicate that systems with a larger magnitude gap would be more dynamically relaxed. Nevertheless, a general study of the dynamical state as a function of the magnitude gap remains to be done.

It is worth noticing that the LFs of the four subsamples---divided for different magnitude-gap bins and both in the regular and relative cases---are reasonably represented by a single Schechter function. Nevertheless, the stacked regular and relative LFs of the 102 systems are not well fitted by a single Schechter function. These differences cannot be due to the method, since only stacked LFs are analyzed. Thus, we interpret this result as another hint that the observed differences in the $M^\ast$ and $\alpha$ parameters of the four subsamples are real and not due to statistic effects.

\subsection{Implications for formation scenarios}

The main results of this work suggest that there is clear dependence on the magnitude gap in the bright part of the LF and a less significant dependence on the magnitude gap in the faint-end slope of the LF.
The characteristic magnitude of the LF can be interpreted as the mean luminosity of the bright galaxy population of a system, when the BGG is excluded from the LF \citep{Cooray2005}. Thus, the observed difference of 1.3 magnitudes in the $M^\ast$ parameter of the regular LF corresponds to a factor $\sim$3 in the mean luminosity. These results are consistent with the most accepted scenario for the formation of the magnitude gap, namely that it developed through dynamical friction, and all the $M^\ast$ galaxies merged in a single, massive central object \citep{DOnghia2004}. However, it is not clear whether the magnitude gap of these systems was formed at high redshift or more recently \citep[][]{Diaz-Gimenez2008}. In fact, \citet{Raouf2014} suggest that $\Delta m_{12}$ alone is not a good age indicator. They claim that there is a trend in age from groups to clusters, where the former are, on average, older than the latter. 

Using their dating method, our sample of fossils would be mainly dominated by young systems. Moreover, \citet{Smith2010} suggest that the formation of the large magnitude gap could depend on both the formation time or the recent infall history of the systems. Thus, a fossil system could also form in a recent epoch, and it could evolve into a regular system in the future, by interacting with other groups/clusters in the same region \citep[see also][]{vonBenda-Beckmann2008}.
Nevertheless, \citet{Deason2013} show that, on average, older and more concentrated halos have larger mass (magnitude) gaps, but the scatter is important because of the transient nature of the satellite population. 

These two results show that the magnitude gap cannot be used as an unambiguous age estimator for individual systems, but that there is a statistical trend in the magnitude gap-age relation. In this work we circumvent this limitation by computing stacked LFs for systems with similar magnitude gaps. For this reason, it is reasonable to assume that the systems in the larger $\Delta m_{12}$ bin are, on average, dynamically older than the others.
But not only age matters. \citet{Sommer-Larsen2006} suggests (using smoothed particle hydrodynamics simulations) that the anisotropy of the orbits can play a role as well. This kind of orbit can bring the infalling population close to the center of the potential well, thus favoring the merging of massive galaxies. They conclude that the more radial the orbits at the time of formation, the more fossil the system will be at the present time.

The observed dependence of the faint-end slope of the LF on the magnitude gap is more intriguing. There is some evidence in the literature showing that some nearby galaxy clusters (e.g., the Coma and the Virgo clusters) contain a small number of dwarf galaxies in the innermost regions \citep[see][]{Aguerri2004,Trujillo2002}. Nevertheless, our results point out that the number of dwarfs in the innermost regions (R $\le$ 0.5 $R_{200}$) depends on the magnitude gap as well. The low masses of dwarf galaxies make them less susceptible to dynamical friction. In addition, the large velocity dispersion in galaxy clusters makes the merging of dwarfs in the recent epoch a very rare event. 

One appealing possibility is that the paucity of dwarfs in systems with large gaps is related to the more radial orbits predicted by \citet{Sommer-Larsen2006}. These eccentric orbits can efficiently bring infalling groups close to the center of the potential well. Once there, not only will the more massive galaxies of these groups merge on a relatively short timescale, but strong tidal forces can efficiently disrupt lower mass halos. Furthermore, if the assembly occurs at relatively early times, the surviving subhalos will spend a significant amount of their history orbiting within rather massive halos---thus increasing the chances of eventual disruption. A similar tidal force-driven disruption mechanism was proposed by \citet{Lopez-Cruz1997} to explain the flattening of the faint-end slope of the LF in the central regions of clusters hosting cD centrals.

%%%%%%%%%%%%%%%%%%%%%%%%%%%%%%%%%%
%%%%%%%%%%%%% SECTION 6%%% %%%%%%%%%%%%
%%%%%%%%%%%%%%%%%%%%%%%%%%%%%%%%%%
\section{Conclusions}
\label{conclusions}

We analyzed a sample of 102 systems with redshift $z \le 0.25$ in order to determine the properties of their LFs. The sample was divided into four subsamples, covering different ranges of $\Delta m_{12}$. In particular, the first subsample included systems with $\Delta m_{12} < 0.5$, the second systems with $0.5 \le \Delta m_{12} < 1.0$, the third systems with $1 \le \Delta m_{12} < 1.5$, and the fourth systems with $\Delta m_{12} \ge 1.5$. The LFs were computed in half the $R_{200}$ radius using a hybrid method, which allowed us to use both photometric and spectroscopic data. Moreover, to better define the differences in the parameters between the four subsamples, we calculated the relative LF. For each galaxy system, the relative LF was obtained  by shifting the LF in magnitude such that the BGG magnitude is zero.

Our results can be summarized as follows:

\begin{itemize}
\item The faint-end slope of the regular stacked LFs of the 102 systems in our sample turns out to be $\alpha=-1.27 \pm 0.11$. The slope of the relative stacked LF is $\alpha=-1.25 \pm 0.09$. The two slopes are in good agreement. These slopes have been obtained by fitting an exponential function and are calculated using the last five points of each LF ($-18 \le M_r \le -16.5$ for the regular LF and $4.5 \le \Delta M_r \le 7$ for the relative LF). We fit an exponential to these stacked LFs because they are not adequately described by either a single or a double Schechter function.
\item In both the regular and relative stacked LFs, if we divide the sample of 102 systems into four subsamples of different $\Delta m_{12}$, the $M^\ast$ values of the Schechter fit change. In particular, the larger the gap, the fainter the $M\ast$, as is expected if the gap is created by dynamical friction. The differences for the subsamples with the smallest and largest gaps are $\sim$ 1.3 magnitude using the regular LF and $\sim$ 2.8 magnitude using the relative LF.
\item The faint-end slope also shows a dependence on the magnitude gap, although the results are less significant.  The Schechter faint-end slopes obtained for the four subsamples follow a trend moving from smaller to larger gaps. The $\alpha$ parameter changes from $\alpha=-1.23$ to $\alpha=-0.78$ in the regular LF, and from $\alpha=-1.26$ to $\alpha=-0.77$ for the relative LF.  We also fit an exponential to the data, finding the same trend in both cases. This is unexpected because the dwarf galaxy population should not be affected by dynamical friction. Other processes, such as more radial orbits or early dwarf galaxy disruption, may play a role. 
\end{itemize}

The uncertainties in the faint-end slopes are mainly due to photometric uncertainties. To improve these results, it is important to study deep spectroscopic LFs for systems with large magnitude gaps. This is part of the future plans of the FOGO project. 

\begin{acknowledgements}
This work has been partially funded by the MINECO (grant AIA2013-43188-P). This article is based on observations made with the Isaac Newton Telescope, Nordic Optical Telescope, and Telescopio Nazionale Galileo operated on the island of La Palma by the Isaac Newton Group, the Nordic Optical Telescope Scientific Association, and the Fundación Galileo Galilei of the INAF (Istituto Nazionale di Astrofisica), respectively, in the Spanish Observatorio del Roque de los Muchachos of the Instituto de Astrofísica de Canarias. E.M.C. acknowledges financial support from Padua University by the grants 60A02-4807/12, 60A02-5857/13, 60A02-5833/14, and CPDA133894. M.G. acknowledges financial support from MIUR PRIN2010-2011. JMA acknowledges support from the European Research Council Starting Grant (SEDmorph; P.I. V. Wild). ED gratefully acknowledges the support of the Alfred P. Sloan Foundation. ED and AK are supported by NASA Grant No NNX13AE97G. J.I.P. and J.M.V. acknowledge financial support from the Spanish MINECO under grant AYA2010-21887-C04-01 and from Junta de Andaluc\'{\i}a Excellence Project PEX2011-FQM7058.

Funding for the Sloan Digital Sky Survey (SDSS) and SDSS-II has been provided by the Alfred P. Sloan Foundation, the Participating Institutions, the National Science Foundation, the U.S. Department of Energy, the National Aeronautics and Space Administration, the Japanese Monbukagakusho, and the Max Planck Society, and the Higher Education Funding Council for England. The SDSS Web site is http://www.sdss.org/.
The SDSS is managed by the Astrophysical Research Consortium (ARC) for the Participating Institutions. The Participating Institutions are the American Museum of Natural History, Astrophysical Institute Potsdam, University of Basel, University of Cambridge, Case Western Reserve University, The University of Chicago, Drexel University, Fermilab, the Institute for Advanced Study, the Japan Participation Group, The Johns Hopkins University, the Joint Institute for Nuclear Astrophysics, the Kavli Institute for Particle Astrophysics and Cosmology, the Korean Scientist Group, the Chinese Academy of Sciences (LAMOST), Los Alamos National Laboratory, the Max-Planck-Institute for Astronomy (MPIA), the Max-Planck-Institute for Astrophysics (MPA), New Mexico State University, Ohio State University, University of Pittsburgh, University of Portsmouth, Princeton University, the United States Naval Observatory, and the University of Washington.
\end{acknowledgements}

\bibliographystyle{aa}
\bibliography{bibliografia}

\begin{thebibliography}{72}
\expandafter\ifx\csname natexlab\endcsname\relax\def\natexlab#1{#1}\fi

\bibitem[{{Abazajian} {et~al.}(2009){Abazajian}, {Adelman-McCarthy},
  {Ag{\"u}eros}, {Allam}, {Allende Prieto}, {An}, {Anderson}, {Anderson},
  {Annis}, {Bahcall}, \& et~al.}]{Abazajian2009}
{Abazajian}, K.~N., {Adelman-McCarthy}, J.~K., {Ag{\"u}eros}, M.~A., {et~al.}
  2009, \apjs, 182, 543

\bibitem[{{Abell} {et~al.}(1989){Abell}, {Corwin}, \& {Olowin}}]{Abell1989}
{Abell}, G.~O., {Corwin}, Jr., H.~G., \& {Olowin}, R.~P. 1989, \apjs, 70, 1

\bibitem[{{Adami} {et~al.}(2012){Adami}, {Jouvel}, {Guennou}, {Le Brun},
  {Durret}, {Clement}, {Clerc}, {Comer{\'o}n}, {Ilbert}, {Lin}, {Russeil}, \&
  {Seemann}}]{Adami2012}
{Adami}, C., {Jouvel}, S., {Guennou}, L., {et~al.} 2012, \aap, 540, A105

\bibitem[{{Adelman-McCarthy} {et~al.}(2007){Adelman-McCarthy}, {Ag{\"u}eros},
  {Allam}, {Anderson}, {Anderson}, {Annis}, {Bahcall}, {Bailer-Jones},
  {Baldry}, {Barentine}, {Beers}, {Belokurov}, {Berlind}, {Bernardi},
  {Blanton}, {Bochanski}, {Boroski}, {Bramich}, {Brewington}, {Brinchmann},
  {Brinkmann}, {Brunner}, {Budav{\'a}ri}, {Carey}, {Carliles}, {Carr},
  {Castander}, {Connolly}, {Cool}, {Cunha}, {Csabai}, {Dalcanton}, {Doi},
  {Eisenstein}, {Evans}, {Evans}, {Fan}, {Finkbeiner}, {Friedman}, {Frieman},
  {Fukugita}, {Gillespie}, {Gilmore}, {Glazebrook}, {Gray}, {Grebel}, {Gunn},
  {de Haas}, {Hall}, {Harvanek}, {Hawley}, {Hayes}, {Heckman}, {Hendry},
  {Hennessy}, {Hindsley}, {Hirata}, {Hogan}, {Hogg}, {Holtzman}, {Ichikawa},
  {Ichikawa}, {Ivezi{\'c}}, {Jester}, {Johnston}, {Jorgensen}, {Juri{\'c}},
  {Kauffmann}, {Kent}, {Kleinman}, {Knapp}, {Kniazev}, {Kron}, {Krzesinski},
  {Kuropatkin}, {Lamb}, {Lampeitl}, {Lee}, {Leger}, {Lima}, {Lin}, {Long},
  {Loveday}, {Lupton}, {Mandelbaum}, {Margon}, {Mart{\'{\i}}nez-Delgado},
  {Matsubara}, {McGehee}, {McKay}, {Meiksin}, {Munn}, {Nakajima}, {Nash},
  {Neilsen}, {Newberg}, {Nichol}, {Nieto-Santisteban}, {Nitta}, {Oyaizu},
  {Okamura}, {Ostriker}, {Padmanabhan}, {Park}, {Peoples}, {Pier}, {Pope},
  {Pourbaix}, {Quinn}, {Raddick}, {Re Fiorentin}, {Richards}, {Richmond},
  {Rix}, {Rockosi}, {Schlegel}, {Schneider}, {Scranton}, {Seljak}, {Sheldon},
  {Shimasaku}, {Silvestri}, {Smith}, {Smol{\v c}i{\'c}}, {Snedden}, {Stebbins},
  {Stoughton}, {Strauss}, {SubbaRao}, {Suto}, {Szalay}, {Szapudi}, {Szkody},
  {Tegmark}, {Thakar}, {Tremonti}, {Tucker}, {Uomoto}, {Vanden Berk},
  {Vandenberg}, {Vidrih}, {Vogeley}, {Voges}, {Vogt}, {Weinberg}, {West},
  {White}, {Wilhite}, {Yanny}, {Yocum}, {York}, {Zehavi}, {Zibetti}, \&
  {Zucker}}]{Adelman-McCarthy2007}
{Adelman-McCarthy}, J.~K., {Ag{\"u}eros}, M.~A., {Allam}, S.~S., {et~al.} 2007,
  \apjs, 172, 634

\bibitem[{{Adelman-McCarthy} {et~al.}(2006){Adelman-McCarthy}, {Ag{\"u}eros},
  {Allam}, {Anderson}, {Anderson}, {Annis}, {Bahcall}, {Baldry}, {Barentine},
  {Berlind}, {Bernardi}, {Blanton}, {Boroski}, {Brewington}, {Brinchmann},
  {Brinkmann}, {Brunner}, {Budav{\'a}ri}, {Carey}, {Carr}, {Castander},
  {Connolly}, {Csabai}, {Czarapata}, {Dalcanton}, {Doi}, {Dong}, {Eisenstein},
  {Evans}, {Fan}, {Finkbeiner}, {Friedman}, {Frieman}, {Fukugita}, {Gillespie},
  {Glazebrook}, {Gray}, {Grebel}, {Gunn}, {Gurbani}, {de Haas}, {Hall},
  {Harris}, {Harvanek}, {Hawley}, {Hayes}, {Hendry}, {Hennessy}, {Hindsley},
  {Hirata}, {Hogan}, {Hogg}, {Holmgren}, {Holtzman}, {Ichikawa}, {Ivezi{\'c}},
  {Jester}, {Johnston}, {Jorgensen}, {Juri{\'c}}, {Kent}, {Kleinman}, {Knapp},
  {Kniazev}, {Kron}, {Krzesinski}, {Kuropatkin}, {Lamb}, {Lampeitl}, {Lee},
  {Leger}, {Lin}, {Long}, {Loveday}, {Lupton}, {Margon},
  {Mart{\'{\i}}nez-Delgado}, {Mandelbaum}, {Matsubara}, {McGehee}, {McKay},
  {Meiksin}, {Munn}, {Nakajima}, {Nash}, {Neilsen}, {Newberg}, {Newman},
  {Nichol}, {Nicinski}, {Nieto-Santisteban}, {Nitta}, {O'Mullane}, {Okamura},
  {Owen}, {Padmanabhan}, {Pauls}, {Peoples}, {Pier}, {Pope}, {Pourbaix},
  {Quinn}, {Richards}, {Richmond}, {Rockosi}, {Schlegel}, {Schneider},
  {Schroeder}, {Scranton}, {Seljak}, {Sheldon}, {Shimasaku}, {Smith}, {Smol{\v
  c}i{\'c}}, {Snedden}, {Stoughton}, {Strauss}, {SubbaRao}, {Szalay},
  {Szapudi}, {Szkody}, {Tegmark}, {Thakar}, {Tucker}, {Uomoto}, {Vanden Berk},
  {Vandenberg}, {Vogeley}, {Voges}, {Vogt}, {Walkowicz}, {Weinberg}, {West},
  {White}, {Xu}, {Yanny}, {Yocum}, {York}, {Zehavi}, {Zibetti}, \&
  {Zucker}}]{Adelman-McCarthy2006}
{Adelman-McCarthy}, J.~K., {Ag{\"u}eros}, M.~A., {Allam}, S.~S., {et~al.} 2006,
  \apjs, 162, 38

\bibitem[{{Aguerri} {et~al.}(2011){Aguerri}, {Girardi}, {Boschin}, {Barrena},
  {M{\'e}ndez-Abreu}, {S{\'a}nchez-Janssen}, {Borgani}, {Castro-Rodriguez},
  {Corsini}, {Del Burgo}, {D'Onghia}, {Iglesias-P{\'a}ramo}, {Napolitano}, \&
  {Vilchez}}]{Aguerri2011}
{Aguerri}, J.~A.~L., {Girardi}, M., {Boschin}, W., {et~al.} 2011, \aap, 527,
  A143

\bibitem[{{Aguerri} {et~al.}(2004){Aguerri}, {Iglesias-Paramo}, {Vilchez}, \&
  {Mu{\~n}oz-Tu{\~n}{\'o}n}}]{Aguerri2004}
{Aguerri}, J.~A.~L., {Iglesias-Paramo}, J., {Vilchez}, J.~M., \&
  {Mu{\~n}oz-Tu{\~n}{\'o}n}, C. 2004, \aj, 127, 1344

\bibitem[{{Aguerri} {et~al.}(2007){Aguerri}, {S{\'a}nchez-Janssen}, \&
  {Mu{\~n}oz-Tu{\~n}{\'o}n}}]{Aguerri2007}
{Aguerri}, J.~A.~L., {S{\'a}nchez-Janssen}, R., \& {Mu{\~n}oz-Tu{\~n}{\'o}n},
  C. 2007, \aap, 471, 17

\bibitem[{{Agulli} {et~al.}(2014){Agulli}, {Aguerri}, {S{\'a}nchez-Janssen},
  {Barrena}, {Diaferio}, {Serra}, \& {M{\'e}ndez-Abreu}}]{Agulli2014}
{Agulli}, I., {Aguerri}, J.~A.~L., {S{\'a}nchez-Janssen}, R., {et~al.} 2014,
  \mnras, 444, L34

\bibitem[{{Barrena} {et~al.}(2012){Barrena}, {Girardi}, {Boschin}, \&
  {Mardirossian}}]{Barrena2012}
{Barrena}, R., {Girardi}, M., {Boschin}, W., \& {Mardirossian}, F. 2012, \aap,
  540, A90

\bibitem[{{Bertin} \& {Arnouts}(1996)}]{Bertin1996}
{Bertin}, E. \& {Arnouts}, S. 1996, \aaps, 117, 393

\bibitem[{{B{\"o}hringer} {et~al.}(2000){B{\"o}hringer}, {Voges}, {Huchra},
  {McLean}, {Giacconi}, {Rosati}, {Burg}, {Mader}, {Schuecker}, {Simi{\c c}},
  {Komossa}, {Reiprich}, {Retzlaff}, \& {Tr{\"u}mper}}]{Boehringer2000}
{B{\"o}hringer}, H., {Voges}, W., {Huchra}, J.~P., {et~al.} 2000, \apjs, 129,
  435

\bibitem[{{Boylan-Kolchin} {et~al.}(2008){Boylan-Kolchin}, {Ma}, \&
  {Quataert}}]{Boylan-Kolchin2008}
{Boylan-Kolchin}, M., {Ma}, C.-P., \& {Quataert}, E. 2008, \mnras, 383, 93

\bibitem[{{Capak} {et~al.}(2004){Capak}, {Cowie}, {Hu}, {Barger}, {Dickinson},
  {Fernandez}, {Giavalisco}, {Komiyama}, {Kretchmer}, {McNally}, {Miyazaki},
  {Okamura}, \& {Stern}}]{Capak2004}
{Capak}, P., {Cowie}, L.~L., {Hu}, E.~M., {et~al.} 2004, \aj, 127, 180

\bibitem[{{Chandrasekhar}(1943)}]{Chandrasekhar1943}
{Chandrasekhar}, S. 1943, \apj, 97, 255

\bibitem[{{Chilingarian} {et~al.}(2010){Chilingarian}, {Melchior}, \&
  {Zolotukhin}}]{Chilingarian2010}
{Chilingarian}, I.~V., {Melchior}, A.-L., \& {Zolotukhin}, I.~Y. 2010, \mnras,
  405, 1409

\bibitem[{{Chilingarian} \& {Zolotukhin}(2012)}]{Chilingarian2012}
{Chilingarian}, I.~V. \& {Zolotukhin}, I.~Y. 2012, \mnras, 419, 1727

\bibitem[{{Colless}(1989)}]{Colless1989}
{Colless}, M. 1989, \mnras, 237, 799

\bibitem[{{Cooray} \& {Milosavljevi{\'c}}(2005)}]{Cooray2005}
{Cooray}, A. \& {Milosavljevi{\'c}}, M. 2005, \apjl, 627, L89

\bibitem[{{Cypriano} {et~al.}(2006){Cypriano}, {Mendes de Oliveira}, \&
  {Sodr{\'e}}}]{Cypriano2006}
{Cypriano}, E.~S., {Mendes de Oliveira}, C.~L., \& {Sodr{\'e}}, Jr., L. 2006,
  \aj, 132, 514

\bibitem[{{Dariush} {et~al.}(2007){Dariush}, {Khosroshahi}, {Ponman}, {Pearce},
  {Raychaudhury}, \& {Hartley}}]{Dariush2007}
{Dariush}, A., {Khosroshahi}, H.~G., {Ponman}, T.~J., {et~al.} 2007, \mnras,
  382, 433

\bibitem[{{Dariush} {et~al.}(2010){Dariush}, {Raychaudhury}, {Ponman},
  {Khosroshahi}, {Benson}, {Bower}, \& {Pearce}}]{Dariush2010}
{Dariush}, A.~A., {Raychaudhury}, S., {Ponman}, T.~J., {et~al.} 2010, \mnras,
  405, 1873

\bibitem[{{de Filippis} {et~al.}(2011){de Filippis}, {Paolillo}, {Longo}, {La
  Barbera}, {de Carvalho}, \& {Gal}}]{deFilippis2011}
{de Filippis}, E., {Paolillo}, M., {Longo}, G., {et~al.} 2011, \mnras, 414,
  2771

\bibitem[{{Deason} {et~al.}(2013){Deason}, {Conroy}, {Wetzel}, \&
  {Tinker}}]{Deason2013}
{Deason}, A.~J., {Conroy}, C., {Wetzel}, A.~R., \& {Tinker}, J.~L. 2013, \apj,
  777, 154

\bibitem[{{D{\'{\i}}az-Gim{\'e}nez} {et~al.}(2008){D{\'{\i}}az-Gim{\'e}nez},
  {Muriel}, \& {Mendes de Oliveira}}]{Diaz-Gimenez2008}
{D{\'{\i}}az-Gim{\'e}nez}, E., {Muriel}, H., \& {Mendes de Oliveira}, C. 2008,
  \aap, 490, 965

\bibitem[{{D'Onghia} \& {Lake}(2004)}]{DOnghia2004}
{D'Onghia}, E. \& {Lake}, G. 2004, \apj, 612, 628

\bibitem[{{D'Onghia} {et~al.}(2005){D'Onghia}, {Sommer-Larsen}, {Romeo},
  {Burkert}, {Pedersen}, {Portinari}, \& {Rasmussen}}]{DOnghia2005}
{D'Onghia}, E., {Sommer-Larsen}, J., {Romeo}, A.~D., {et~al.} 2005, \apjl, 630,
  L109

\bibitem[{{Eigenthaler} \& {Zeilinger}(2013)}]{Eigenthaler2013}
{Eigenthaler}, P. \& {Zeilinger}, W.~W. 2013, \aap, 553, A99

\bibitem[{{Girardi} {et~al.}(2014){Girardi}, {Aguerri}, {De Grandi},
  {D'Onghia}, {Barrena}, {Boschin}, {M{\'e}ndez-Abreu}, {S{\'a}nchez-Janssen},
  {Zarattini}, {Biviano}, {Castro-Rodriguez}, {Corsini}, {del Burgo},
  {Iglesias-P{\'a}ramo}, \& {Vilchez}}]{Girardi2014}
{Girardi}, M., {Aguerri}, J.~A.~L., {De Grandi}, S., {et~al.} 2014, \aap, 565,
  A115

\bibitem[{{Gozaliasl} {et~al.}(2014){Gozaliasl}, {Khosroshahi}, {Dariush},
  {Finoguenov}, {Jassur}, \& {Molaeinajad}}]{Gozaliasl2014}
{Gozaliasl}, G., {Khosroshahi}, H.~G., {Dariush}, A.~A., {et~al.} 2014, ArXiv
  e-prints

\bibitem[{{Harrison} {et~al.}(2012){Harrison}, {Miller}, {Richards},
  {Lloyd-Davies}, {Hoyle}, {Romer}, {Mehrtens}, {Hilton}, {Stott}, {Capozzi},
  {Collins}, {Deadman}, {Liddle}, {Sahl{\'e}n}, {Stanford}, \&
  {Viana}}]{Harrison2012}
{Harrison}, C.~D., {Miller}, C.~J., {Richards}, J.~W., {et~al.} 2012, \apj,
  752, 12

\bibitem[{{Hearin} {et~al.}(2013){Hearin}, {Zentner}, {Berlind}, \&
  {Newman}}]{Hearin2013}
{Hearin}, A.~P., {Zentner}, A.~R., {Berlind}, A.~A., \& {Newman}, J.~A. 2013,
  \mnras, 433, 659

\bibitem[{{Huang} {et~al.}(1997){Huang}, {Cowie}, {Gardner}, {Hu}, {Songaila},
  {Wainscoat}, \& {R.~J.}}]{huang1997}
{Huang}, J.-S., {Cowie}, L.~L., {Gardner}, J.~P., {et~al.} 1997, \apj, 476, 12

\bibitem[{{Huang} {et~al.}(2001){Huang}, {Thompson}, {K{\"u}mmel},
  {Meisenheimer}, {Wolf}, {Beckwith}, {Fockenbrock}, {Fried}, {Hippelein}, {von
  Kuhlmann}, {Phleps}, {R{\"o}ser}, \& {Thommes}}]{Huang2001}
{Huang}, J.-S., {Thompson}, D., {K{\"u}mmel}, M.~W., {et~al.} 2001, \aap, 368,
  787

\bibitem[{{Iglesias-P{\'a}ramo} {et~al.}(2003){Iglesias-P{\'a}ramo}, {Boselli},
  {Gavazzi}, {Cortese}, \& {V{\'{\i}}lchez}}]{Iglesias-Paramo2003}
{Iglesias-P{\'a}ramo}, J., {Boselli}, A., {Gavazzi}, G., {Cortese}, L., \&
  {V{\'{\i}}lchez}, J.~M. 2003, \aap, 397, 421

\bibitem[{{Johnston}(2011)}]{Johnston2011}
{Johnston}, R. 2011, \aapr, 19, 41

\bibitem[{{Jones} {et~al.}(2003){Jones}, {Ponman}, {Horton}, {Babul},
  {Ebeling}, \& {Burke}}]{Jones2003}
{Jones}, L.~R., {Ponman}, T.~J., {Horton}, A., {et~al.} 2003, \mnras, 343, 627

\bibitem[{{Khosroshahi} {et~al.}(2014){Khosroshahi}, {Gozaliasl}, {Rasmussen},
  {Molaeinezhad}, {Ponman}, {Dariush}, \& {Sanderson}}]{Khosroshahi2014}
{Khosroshahi}, H.~G., {Gozaliasl}, G., {Rasmussen}, J., {et~al.} 2014, \mnras,
  443, 318

\bibitem[{{Khosroshahi} {et~al.}(2006){Khosroshahi}, {Maughan}, {Ponman}, \&
  {Jones}}]{Khosroshahi2006}
{Khosroshahi}, H.~G., {Maughan}, B.~J., {Ponman}, T.~J., \& {Jones}, L.~R.
  2006, \mnras, 369, 1211

\bibitem[{{Khosroshahi} {et~al.}(2007){Khosroshahi}, {Ponman}, \&
  {Jones}}]{Khosroshahi2007}
{Khosroshahi}, H.~G., {Ponman}, T.~J., \& {Jones}, L.~R. 2007, \mnras, 377, 595

\bibitem[{{La Barbera} {et~al.}(2009){La Barbera}, {de Carvalho}, {de la Rosa},
  {Sorrentino}, {Gal}, \& {Kohl-Moreira}}]{LaBarbera2009}
{La Barbera}, F., {de Carvalho}, R.~R., {de la Rosa}, I.~G., {et~al.} 2009,
  \aj, 137, 3942

\bibitem[{{Lieder} {et~al.}(2013){Lieder}, {Mieske}, {S{\'a}nchez-Janssen},
  {Hilker}, {Lisker}, \& {Tanaka}}]{Lieder2013}
{Lieder}, S., {Mieske}, S., {S{\'a}nchez-Janssen}, R., {et~al.} 2013, \aap,
  559, A76

\bibitem[{{L{\'o}pez-Cruz} {et~al.}(1997){L{\'o}pez-Cruz}, {Yee}, {Brown},
  {Jones}, \& {Forman}}]{Lopez-Cruz1997}
{L{\'o}pez-Cruz}, O., {Yee}, H.~K.~C., {Brown}, J.~P., {Jones}, C., \&
  {Forman}, W. 1997, \apjl, 475, L97

\bibitem[{{Mendes de Oliveira} {et~al.}(2006){Mendes de Oliveira}, {Cypriano},
  \& {Sodr{\'e}}}]{MendesdeOliveira2006}
{Mendes de Oliveira}, C.~L., {Cypriano}, E.~S., \& {Sodr{\'e}}, Jr., L. 2006,
  \aj, 131, 158

\bibitem[{{M{\'e}ndez-Abreu} {et~al.}(2012){M{\'e}ndez-Abreu}, {Aguerri},
  {Barrena}, {S{\'a}nchez-Janssen}, {Boschin}, {Castro-Rodriguez}, {Corsini},
  {Del Burgo}, {D'Onghia}, {Girardi}, {Iglesias-P{\'a}ramo}, {Napolitano},
  {Vilchez}, \& {Zarattini}}]{Mendez-Abreu2012}
{M{\'e}ndez-Abreu}, J., {Aguerri}, J.~A.~L., {Barrena}, R., {et~al.} 2012,
  \aap, 537, A25

\bibitem[{{Metcalfe} {et~al.}(2001){Metcalfe}, {Shanks}, {Campos}, {McCracken},
  \& {Fong}}]{Metcalfe2001}
{Metcalfe}, N., {Shanks}, T., {Campos}, A., {McCracken}, H.~J., \& {Fong}, R.
  2001, \mnras, 323, 795

\bibitem[{{Mulchaey} \& {Zabludoff}(1999)}]{Mulchaey1999}
{Mulchaey}, J.~S. \& {Zabludoff}, A.~I. 1999, \apj, 514, 133

\bibitem[{{Munari} {et~al.}(2013){Munari}, {Biviano}, {Borgani}, {Murante}, \&
  {Fabjan}}]{Munari2013}
{Munari}, E., {Biviano}, A., {Borgani}, S., {Murante}, G., \& {Fabjan}, D.
  2013, \mnras, 430, 2638

\bibitem[{{Ponman} {et~al.}(1994){Ponman}, {Allan}, {Jones}, {Merrifield},
  {McHardy}, {Lehto}, \& {Luppino}}]{Ponman1994}
{Ponman}, T.~J., {Allan}, D.~J., {Jones}, L.~R., {et~al.} 1994, \nat, 369, 462

\bibitem[{{Popesso} {et~al.}(2005){Popesso}, {B{\"o}hringer}, {Romaniello}, \&
  {Voges}}]{Popesso2005}
{Popesso}, P., {B{\"o}hringer}, H., {Romaniello}, M., \& {Voges}, W. 2005,
  \aap, 433, 415

\bibitem[{{Proctor} {et~al.}(2011){Proctor}, {de Oliveira}, {Dupke}, {de
  Oliveira}, {Cypriano}, {Miller}, \& {Rykoff}}]{Proctor2011}
{Proctor}, R.~N., {de Oliveira}, C.~M., {Dupke}, R., {et~al.} 2011, \mnras,
  418, 2054

\bibitem[{{Raouf} {et~al.}(2014){Raouf}, {Khosroshahi}, {Ponman}, {Dariush},
  {Molaeinezhad}, \& {Tavasoli}}]{Raouf2014}
{Raouf}, M., {Khosroshahi}, H.~G., {Ponman}, T.~J., {et~al.} 2014, \mnras, 442,
  1578

\bibitem[{{Rines} \& {Geller}(2008)}]{Rines2008}
{Rines}, K. \& {Geller}, M.~J. 2008, \aj, 135, 1837

\bibitem[{{Rykoff} {et~al.}(2008){Rykoff}, {Evrard}, {McKay}, {Becker},
  {Johnston}, {Koester}, {Nord}, {Rozo}, {Sheldon}, {Stanek}, \&
  {Wechsler}}]{Rykoff2008}
{Rykoff}, E.~S., {Evrard}, A.~E., {McKay}, T.~A., {et~al.} 2008, \mnras, 387,
  L28

\bibitem[{{S{\'a}nchez-Janssen} {et~al.}(2008){S{\'a}nchez-Janssen}, {Aguerri},
  \& {Mu{\~n}oz-Tu{\~n}{\'o}n}}]{Sanchez-Janssen2008}
{S{\'a}nchez-Janssen}, R., {Aguerri}, J.~A.~L., \& {Mu{\~n}oz-Tu{\~n}{\'o}n},
  C. 2008, \apjl, 679, L77

\bibitem[{{Santos} {et~al.}(2007){Santos}, {Mendes de Oliveira}, \&
  {Sodr{\'e}}}]{Santos2007}
{Santos}, W.~A., {Mendes de Oliveira}, C., \& {Sodr{\'e}}, Jr., L. 2007, \aj,
  134, 1551

\bibitem[{{Schechter}(1976)}]{Schechter1976}
{Schechter}, P. 1976, \apj, 203, 297

\bibitem[{{Smith} {et~al.}(2010){Smith}, {Khosroshahi}, {Dariush}, {Sanderson},
  {Ponman}, {Stott}, {Haines}, {Egami}, \& {Stark}}]{Smith2010}
{Smith}, G.~P., {Khosroshahi}, H.~G., {Dariush}, A., {et~al.} 2010, \mnras,
  409, 169

\bibitem[{{Sommer-Larsen}(2006)}]{Sommer-Larsen2006}
{Sommer-Larsen}, J. 2006, \mnras, 369, 958

\bibitem[{{Stoughton} {et~al.}(2002){Stoughton}, {Lupton}, {Bernardi},
  {Blanton}, {Burles}, {Castander}, {Connolly}, {Eisenstein}, {Frieman},
  {Hennessy}, {Hindsley}, {Ivezi{\'c}}, {Kent}, {Kunszt}, {Lee}, {Meiksin},
  {Munn}, {Newberg}, {Nichol}, {Nicinski}, {Pier}, {Richards}, {Richmond},
  {Schlegel}, {Smith}, {Strauss}, {SubbaRao}, {Szalay}, {Thakar}, {Tucker},
  {Vanden Berk}, {Yanny}, {Adelman}, {Anderson}, {Anderson}, {Annis},
  {Bahcall}, {Bakken}, {Bartelmann}, {Bastian}, {Bauer}, {Berman},
  {B{\"o}hringer}, {Boroski}, {Bracker}, {Briegel}, {Briggs}, {Brinkmann},
  {Brunner}, {Carey}, {Carr}, {Chen}, {Christian}, {Colestock}, {Crocker},
  {Csabai}, {Czarapata}, {Dalcanton}, {Davidsen}, {Davis}, {Dehnen},
  {Dodelson}, {Doi}, {Dombeck}, {Donahue}, {Ellman}, {Elms}, {Evans}, {Eyer},
  {Fan}, {Federwitz}, {Friedman}, {Fukugita}, {Gal}, {Gillespie}, {Glazebrook},
  {Gray}, {Grebel}, {Greenawalt}, {Greene}, {Gunn}, {de Haas}, {Haiman},
  {Haldeman}, {Hall}, {Hamabe}, {Hansen}, {Harris}, {Harris}, {Harvanek},
  {Hawley}, {Hayes}, {Heckman}, {Helmi}, {Henden}, {Hogan}, {Hogg}, {Holmgren},
  {Holtzman}, {Huang}, {Hull}, {Ichikawa}, {Ichikawa}, {Johnston}, {Kauffmann},
  {Kim}, {Kimball}, {Kinney}, {Klaene}, {Kleinman}, {Klypin}, {Knapp},
  {Korienek}, {Krolik}, {Kron}, {Krzesi{\'n}ski}, {Lamb}, {Leger},
  {Limmongkol}, {Lindenmeyer}, {Long}, {Loomis}, {Loveday}, {MacKinnon},
  {Mannery}, {Mantsch}, {Margon}, {McGehee}, {McKay}, {McLean}, {Menou},
  {Merelli}, {Mo}, {Monet}, {Nakamura}, {Narayanan}, {Nash}, {Neilsen},
  {Newman}, {Nitta}, {Odenkirchen}, {Okada}, {Okamura}, {Ostriker}, {Owen},
  {Pauls}, {Peoples}, {Peterson}, {Petravick}, {Pope}, {Pordes}, {Postman},
  {Prosapio}, {Quinn}, {Rechenmacher}, {Rivetta}, {Rix}, {Rockosi}, {Rosner},
  {Ruthmansdorfer}, {Sandford}, {Schneider}, {Scranton}, {Sekiguchi}, {Sergey},
  {Sheth}, {Shimasaku}, {Smee}, {Snedden}, {Stebbins}, {Stubbs}, {Szapudi},
  {Szkody}, {Szokoly}, {Tabachnik}, {Tsvetanov}, {Uomoto}, {Vogeley}, {Voges},
  {Waddell}, {Walterbos}, {Wang}, {Watanabe}, {Weinberg}, {White}, {White},
  {Wilhite}, {Wolfe}, {Yasuda}, {York}, {Zehavi}, \& {Zheng}}]{Stoughton2002}
{Stoughton}, C., {Lupton}, R.~H., {Bernardi}, M., {et~al.} 2002, \aj, 123, 485

\bibitem[{{Trentham} \& {Hodgkin}(2002)}]{Trentham2002}
{Trentham}, N. \& {Hodgkin}, S. 2002, \mnras, 333, 423

\bibitem[{{Trujillo} {et~al.}(2002){Trujillo}, {Aguerri}, {Guti{\'e}rrez},
  {Caon}, \& {Cepa}}]{Trujillo2002}
{Trujillo}, I., {Aguerri}, J.~A.~L., {Guti{\'e}rrez}, C.~M., {Caon}, N., \&
  {Cepa}, J. 2002, \apjl, 573, L9

\bibitem[{{Voevodkin} {et~al.}(2010){Voevodkin}, {Borozdin}, {Heitmann},
  {Habib}, {Vikhlinin}, {Mescheryakov}, {Hornstrup}, \&
  {Burenin}}]{Voevodkin2010}
{Voevodkin}, A., {Borozdin}, K., {Heitmann}, K., {et~al.} 2010, \apj, 708, 1376

\bibitem[{{Voges} {et~al.}(1999){Voges}, {Aschenbach}, {Boller},
  {Br{\"a}uninger}, {Briel}, {Burkert}, {Dennerl}, {Englhauser}, {Gruber},
  {Haberl}, {Hartner}, {Hasinger}, {K{\"u}rster}, {Pfeffermann}, {Pietsch},
  {Predehl}, {Rosso}, {Schmitt}, {Tr{\"u}mper}, \& {Zimmermann}}]{Voges1999}
{Voges}, W., {Aschenbach}, B., {Boller}, T., {et~al.} 1999, \aap, 349, 389

\bibitem[{{von Benda-Beckmann} {et~al.}(2008){von Benda-Beckmann}, {D'Onghia},
  {Gottl{\"o}ber}, {Hoeft}, {Khalatyan}, {Klypin}, \&
  {M{\"u}ller}}]{vonBenda-Beckmann2008}
{von Benda-Beckmann}, A.~M., {D'Onghia}, E., {Gottl{\"o}ber}, S., {et~al.}
  2008, \mnras, 386, 2345

\bibitem[{{Weinmann} {et~al.}(2011){Weinmann}, {Lisker}, {Guo}, {Meyer}, \&
  {Janz}}]{Weinmann2011}
{Weinmann}, S.~M., {Lisker}, T., {Guo}, Q., {Meyer}, H.~T., \& {Janz}, J. 2011,
  \mnras, 416, 1197

\bibitem[{{Wen} \& {Han}(2013)}]{Wen2013}
{Wen}, Z.~L. \& {Han}, J.~L. 2013, \mnras, 436, 275

\bibitem[{{Wen} \& {Han}(2015)}]{Wen2015}
{Wen}, Z.~L. \& {Han}, J.~L. 2015, \mnras, 448, 2

\bibitem[{{Yasuda} {et~al.}(2001){Yasuda}, {Fukugita}, {Narayanan}, {Lupton},
  {Strateva}, {Strauss}, {Ivezi{\'c}}, {Kim}, {Hogg}, {Weinberg}, {Shimasaku},
  {Loveday}, {Annis}, {Bahcall}, {Blanton}, {Brinkmann}, {Brunner}, {Connolly},
  {Csabai}, {Doi}, {Hamabe}, {Ichikawa}, {Ichikawa}, {Johnston}, {Knapp},
  {Kunszt}, {Lamb}, {McKay}, {Munn}, {Nichol}, {Okamura}, {Schneider},
  {Szokoly}, {Vogeley}, {Watanabe}, \& {York}}]{Yasuda2001}
{Yasuda}, N., {Fukugita}, M., {Narayanan}, V.~K., {et~al.} 2001, \aj, 122, 1104

\bibitem[{{Zandivarez} \& {Mart{\'{\i}}nez}(2011)}]{Zandivarez2011}
{Zandivarez}, A. \& {Mart{\'{\i}}nez}, H.~J. 2011, \mnras, 415, 2553

\bibitem[{{Zarattini} {et~al.}(2014){Zarattini}, {Barrena}, {Girardi},
  {Castro-Rodriguez}, {Boschin}, {Aguerri}, {M{\'e}ndez-Abreu},
  {S{\'a}nchez-Janssen}, {Catal{\'a}n-Torrecilla}, {Corsini}, {del Burgo},
  {D'Onghia}, {Herrera-Ruiz}, {Iglesias-P{\'a}ramo}, {Jimenez Bailon}, {Lozada
  Muoz}, {Napolitano}, \& {Vilchez}}]{Zarattini2014}
{Zarattini}, S., {Barrena}, R., {Girardi}, M., {et~al.} 2014, \aap, 565, A116

\bibitem[{{Zwicky} {et~al.}(1961){Zwicky}, {Herzog}, {Wild}, {Karpowicz}, \&
  {Kowal}}]{Zwicky1961}
{Zwicky}, F., {Herzog}, E., {Wild}, P., {Karpowicz}, M., \& {Kowal}, C.~T.
  1961, {Catalogue of galaxies and of clusters of galaxies, Vol. I} (Pasadena:
  California Institute of Technology (CIT))

\end{thebibliography}

%__________________________________________________________________
\end{document}